\documentclass[acmsmall]{acmart}
\usepackage[normalem]{ulem}
 \usepackage{amsmath}
  \usepackage{multirow}
\usepackage{epstopdf}
  \usepackage{graphicx}
\usepackage{appendix}
\usepackage[normalem]{ulem}
\useunder{\uline}{\ul}{}
 \usepackage{enumitem}
 \useunder{\uline}{\ul}{}
\AtBeginDocument{%
  \providecommand\BibTeX{{%
    \normalfont B\kern-0.5em{\scshape i\kern-0.25em b}\kern-0.8em\TeX}}}
    
\usepackage{subfigure}

\newcommand{\chenc}[1]{\textcolor{black}{#1}}

\setcopyright{rightsretained}

\usepackage{etoolbox}
  
\author{Guanyu Lin}
\affiliation{%
  \institution{Carnegie Mellon University}  \country{United States}}
 
\email{guanyul@andrew.cmu.edu}

\author{Chen Gao}
\affiliation{%
  \institution{Tsinghua University}  \country{China} }
\email{chgao96@gmail.com}

\author{Yinfeng Li}
\affiliation{%
  \institution{Tsinghua University}\country{China}}
  \email{liyf19@mails.tsinghua.edu.cn}
  
\author{Yu Zheng}
\affiliation{%
  \institution{Tsinghua University}\country{China}}
  \email{y-zheng19@mails.tsinghua.edu.cn}
\author{Zhiheng Li}
\affiliation{%
  \institution{Tsinghua University}\country{China}
  }
  \email{zhhli@mail.tsinghua.edu.cn}
\author{Depeng Jin}
\affiliation{%
  \institution{Tsinghua University}\country{China}}
  \email{jindp@tsinghua.edu.cn}

\author{Dong Li}
\affiliation{%
  \institution{Huawei Noah’s Ark Lab}\country{China}}
  \email{lidong106@huawei.com}

\author{Jianye Hao}
\affiliation{%
  \institution{Huawei Noah’s Ark Lab}\country{China}}
  \email{haojianye@huawei.com}

\author{Yong Li}
\affiliation{%
  \institution{Tsinghua University}\country{China}}
  \email{liyong07@tsinghua.edu.cn}
\begin{document}
\fancyhead{}

\title{Mutual Harmony: Sequential Recommendation with Dual Contrastive Network}

\begin{abstract}
With the outbreak of today's streaming data, the sequential recommendation is a promising solution to achieve time-aware personalized modeling. It aims to infer the next interacted item of a given user based on the historical item sequence. Some recent works tend to improve the sequential recommendation via random masking on the historical item so as to generate self-supervised signals. But such \chenc{approaches} will indeed result in sparser item sequence and unreliable signals. Besides, the existing sequential recommendation \chenc{models are} only user-centric, \textit{i.e.,} based on the historical items by chronological order to predict the probability of candidate items, which ignores whether the items from a provider can be successfully recommended. Such user-centric recommendation will make it impossible for the provider to expose their new items, failing to consider the accordant interactions between user and item dimensions.

In this paper, we propose a novel \textbf{D}ual \textbf{C}ontrastive \textbf{N}etwork (DCN) to achieve mutual harmony between user and item provider, generating ground-truth self-supervised signals for sequential recommendation by auxiliary user-sequence from an item-centric dimension. Specifically, we propose dual representation contrastive learning to refine the representation learning by minimizing the Euclidean distance between the representations of a given user/item and historical items/users of them. 
Before the second contrastive learning module, we perform the next user prediction to capture the trends of items preferred by certain types of users and provide personalized exploration opportunities for item providers. Finally, we further propose dual interest contrastive learning to self-supervise the dynamic interest from the next item/user prediction and static interest of matching probability. 
Experiments on four benchmark datasets verify the effectiveness of our proposed method. Further ablation study also illustrates the boosting effect of the proposed components upon different sequential models. \chenc{We have released the source code at: \url{https://github.com/tsinghua-fib-lab/DCN}}.


\end{abstract}



\maketitle
\section{Introduction}\label{sec:intro}
\begin{figure}[t!]
		\centering
		\begin{tabular}{c|c}
		\includegraphics[width=.45\linewidth]{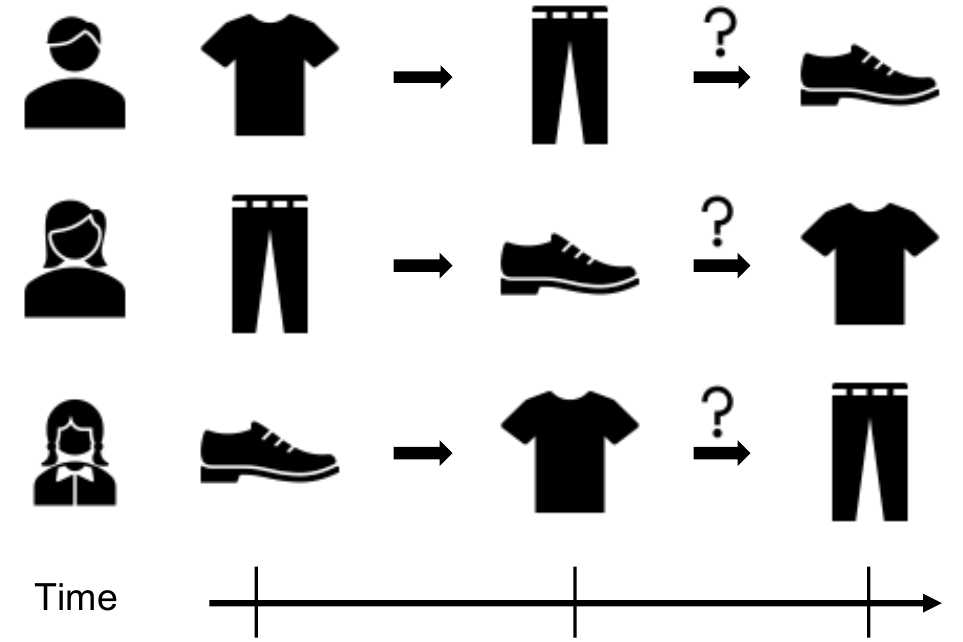}  & \includegraphics[width=.45\linewidth]{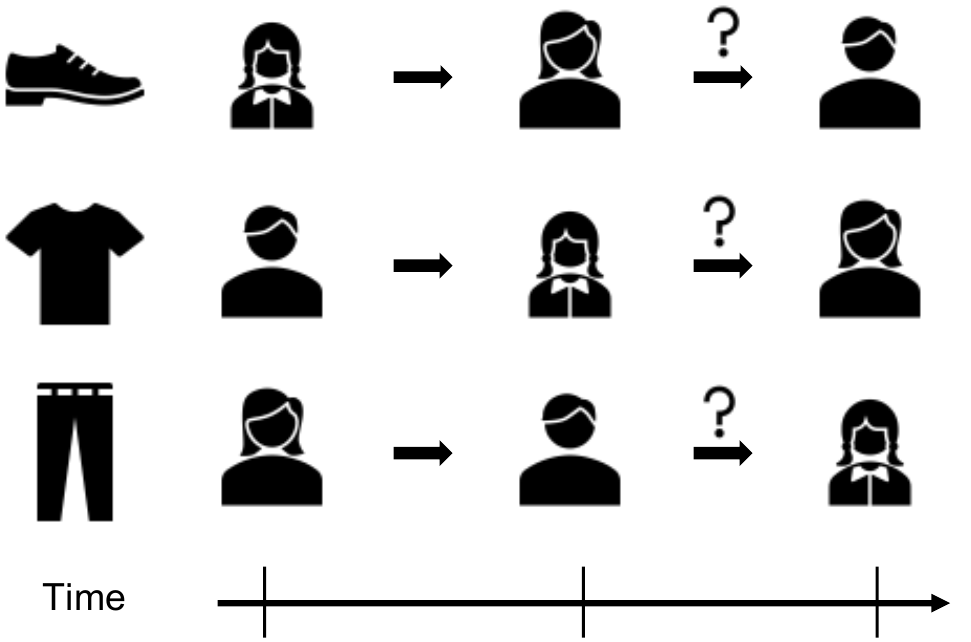} \\
		(a) Item sequence & (b) User sequence
		\end{tabular}
\caption{Illustration of item and user sequence. (a) Traditional sequential recommendation based on item sequence (next item prediction) aims to predict the probability of the target item that the target user will click. (b) Indeed, each interaction of item sequence inherently includes a user sequence, which can be exploited to predict the next user and is ignored by existing work.}	\label{fig:item_user_sequence}
\end{figure} 
Widespread in today's big data applications, recommender systems that vastly facilitate users’ soaking or purchasing behavior are vital for an online platform's business success. Popularly studied in both industrial and academic recommendation, sequential recommendation~\cite{SRs} is a promising approach to further model the dynamic interest of users.


As shown in Figure~\ref{fig:item_user_sequence}(a), traditional sequential recommendation is always user-centric, \textit{i.e.,} predicting a user's next interacted item based on his/her historical item sequence. However, as shown in Figure~\ref{fig:item_user_sequence}(b), there is another user sequence from the item-centric dimension, which is ignored by almost all existing sequential recommendations. Ideally, a trustworthy recommender systems should achieve mutual harmony on both user and item-centric dimensions.
Some advanced deep learning methods such as recurrent neural network~\cite{mikolov2010recurrent, GRU, LSTM}, convolution neural network~\cite{CNN} and attention network~\cite{vaswani2017attention} have been applied in modern sequential recommendation~\cite{GRU4REC, DIEN, Caser, SASRec} to capture the high-order transition between the historical items.
Despite the achievements of existing user-centric sequential recommendation models, they suffer from data sparsity issue~\cite{wang_sequential_2019, wang_counterfactual_2021}. To improve sequential recommendation models, some recent works~\cite{sun2019bert4rec, zhou2020s3} have attempted to exploit self-supervised learning~\cite{hjelm2018learning}. \chenc{The existing self-supervised sequential recommendation mostly generates augmented data via dropout strategy. Such strategy indeed may lead to sparser sequential data and unreliable signals, especially when the user behavior sequence is short.} To explore self-supervised signals from a new perspective, a session-based recommendation work~\cite{xia2021self} exploits the inherent self-supervised signals from the session view and item view of one sequence. However, there is no session view for the traditional sequential recommendation. \chenc{Here, session view means the session sequence of the user in the session-based recommendation. For example, if a user opens the app in the morning and afternoon, these the morning session and afternoon session can be linked together as a session sequence.}
But fortunately, \chenc{as the user sequence and item sequence are generated from the same interaction event}, user sequence is an inherent self-supervised signal for item sequence. Thus it is an elegant way to improve the existing item sequential recommendation by incorporating user sequence based on contrastive learning. It not only can address the bottleneck of existing self-supervised sequential recommendation but also can extend it to further consider item-centric dimension, hitting two birds with one stone.


Indeed, user-centric item recommendation fails to recommend the potential users for the item providers. Besides, user-centric item recommendation tends to be overfitting on the exposed data and limited in the popular items, which makes the exposed items more and more popular~\cite{liu2019spiral, schnabel2016recommendations, yang2018unbiased}, keeping new items from being exposed. These new items may be potentially popular and will be preferred by certain types of users in the future.
Actually, if we provide some opportunities for the item providers to expose their new and \chenc{cold} items to potential users, the popular bias \chenc{with over-recommendation of old popular items} can be avoided. Therefore, the interactions between users and item providers are accordant and simultaneously meeting these two dimensions can promote each other.

Both users and item providers are two significant dimensions of recommender systems. Thus we should not only provide the opportunities for the users to actively consume the items but also for the item providers to actively expose their items~\cite{Fairness, SocialWelfare, MatchingMarkets}.
   That is to say, it is also important to take care of the item-centric dimension and capture the dynamic trend of user sequence to perform user recommendations for item providers. \chenc{Indeed, such item-centric dimension tends to find users that are more likely to interact with particular items, which is also called “audience segmentation” in existing works~\cite{RTB, yuan2014survey} for real-time bidding. Specifically, these works propose audience segmentation to promote the conversion of advertisements. Different from these real-time bidding works~\cite{RTB, yuan2014survey}, we propose user recommendation as a novel solution to self-supervise with the item recommendation.}
   To extract the potential users for the new items, it is necessary to recommend users for the item providers. Note that we do not ignore the possibility of disturbance of item providers to the users, but we suppose a few users recommended to the item providers will be helpful to the exposure of new items.
To achieve effective modeling from both user-centric and item-centric dimensions, there is also a work~\cite{ma2021bridging} deriving Recurrent Intensity Models (RIMs) as unifying extensions from item-centric session-based recommendation while we further incorporate the user sequence into the sequential recommendation.
It is challenging to introduce user sequence. If we simply combine the sequential embeddings of item sequence and user sequence, it may be difficult for the model to distinguish them. Thus we need to design some contrastive learning methods to extract effective signals between user and item dimensions.

In this paper, we propose a novel solution named DCN (short for \textbf{D}ual \textbf{C}ontrastive \textbf{N}etwork for Sequential Recommendation) based on auxiliary user sequence. In contrast to item sequence that reflects the dynamic interest of the given user, user sequence reflects the dynamic trend of a given item preferred by certain types of users. Thus, user sequence serves as a complementary role to a traditional sequential recommendation based on item sequence.
%
Specifically, we first encode the user/item sequence for a given target item/user based on the user/item encoder, respectively.
Then we treat the output of the user/item encoder as the representation of the given target item/user and perform contrastive learning to minimize the distance between it and the target item/user embedding, extracting self-supervised signals for representation learning.
Finally, to capture the dynamic interest and static interest simultaneously, we perform contrastive learning on the next item prediction with the target user-item interaction prediction, as well as the next user prediction. 

Note that existing sequential recommendation is often user-centric and evaluated on item recommendation metric, failing to consider the item-centric dimension. If simply applying the next user prediction from an item-centric dimension, it will perform poorly in the evaluation metric and collected data by user-centric recommenders. Besides, if exploiting item and user sequence to predict the next item and user independently, they will fail to augment each other. However, it will boost the performance if taking the user sequence as an auxiliary training signal. Our method improves existing sequential recommendations from a totally new perspective. We evaluate our proposed method not only on the traditional user-centric evaluation metric, i.e., item recommendation perspective, but also on a new item-centric evaluation metric, i.e., user recommendation perspective. To summarize, the contribution of this paper is as below.
 \begin{itemize}[leftmargin=*] 
 	\item To the best of our knowledge, we approach the problem of sequential recommendation from a more systematic view, \textit{i.e.}, considering the accordant 
 interactions between user and item dimensions simultaneously.
 		\item We perform contrastive learning from both representation-aspect and interest-aspect, respectively, to refine the user-item representation learning and capture both the static and dynamic interest.
	\item We conduct experiments on four benchmark datasets where the results on both user-centric and item-centric evaluation metrics outperform the state-of-the-art models. 
	Further experimental results of the ablation study will confirm the effectiveness of our designed self-supervision.
 \end{itemize}
 
An early version of this work has been published as a 6-page short paper in SIGIR 2022~\cite{dcn22}, compared with which
we have the following new contributions: (1) We perform a representation transformation layer on the sequential embeddings, transforming the item sequence embeddings to user representation and the user sequence embeddings to item representation before performing dual representation contrastive learning; (2) We approach the problem of sequential recommendation from a more systematic view, further evaluating the proposed method from item-centric dimension, instead of solely user-centric dimension; (3) We conduct experiments on additional two benchmark datasets where the results on both item and user recommendations outperform the state-of-the-art models; (4) Further experimental results of the ablation study on additional datasets also well confirm the effectiveness of our designed self-supervision; (5) We also further perform a hyper-parameter study on the regularization parameter of contrastive loss.

 	


The remaining part of this paper is organized as follows. Firstly, we formulate the problem user\&item-centric recommendation in Section~\ref{sec:proDef}, comparing it with user-centric and item-centric recommendations, respectively. Then we illustrate the proposed method from the encoding layer, dual representation contrastive layer, mixed prediction layer, and dual interest contrastive layer in Section~\ref{sec:method}. In Section~\ref{sec:exp}, we conduct experiments on both user-centric and item-centric evaluation to compare with the existing sequential recommendation baselines when further ablation and backbone studies illustrate the effectiveness of our proposed components. Finally, we draw a conclusion before prospecting future work.


\section{{Related Work}}

\subsection{\textbf{\chenc{Sequential Recommendation}}}

One related work of our solution is Sequential Recommendation~\cite{SRs}, which aims to predict the probability of the next interacted item based on the historical item sequence of the given target user. Widely researched in the academic institute, the sequential recommendation is also widely applied in industrial platforms~\cite{acf17, latentCross18, SIM20, sam22}. Markov chain is the most initial sequential model applied in the sequential recommendation, namely FPMC~\cite{rendle2010factorizing}. Subsequently, some sprouted deep learning models such as recurrent neural network~\cite{GRU, LSTM}, convolution neural network~\cite{CNN} and attention network~\cite{vaswani2017attention} are then applied in recommendation ~\cite{GRU4REC, DIEN, Caser, SASRec, DIN} to achieve better generalization. To further capture the short-term and long-term interest simultaneously, matrix factorization~\cite{koren2009matrix} is also adopted to achieve fancy performance~\cite{SLIREC, zhao2018plastic}. RIMs~\cite{ma2021bridging} connects personalized item recommendation and marketing user recommendation by dual decomposition to trade-off between recommendation relevance and item diversity.
However, existing works of sequential recommendation only consider the item sequence side without further considering the user sequence side. Indeed, item sequence reflects the dynamic interest of given user preferring items while user sequence reflects the dynamic trend of given item preferred by users.

\subsection{\textbf{\chenc{Self-supervised Learning}}}

Recently, self-supervised learning~\cite{hjelm2018learning} is a popular machine learning method, which supervises the model via generated labels and promotes optimization via the generated self-supervised signals. In general, self-supervised learning often exploits the learned representation to generate the supervised signals, which can be applied in and promote the downstream task. 
 Indeed, self-supervised learning has already achieved decent performance in the fields of natural language processing~\cite{devlin2018bert} and computer vision~\cite{bachman2019learning}, based on the random dropout upon raw data information such as masking the sentence or rotating/clipping the image. Advanced self-supervised learning also focuses on graph representation learning and has achieved decent performance~\cite{hu2019strategies, velickovic2019deep, wu2021self}, where DGI~\cite{velickovic2019deep} maximizes the mutual information between the local part (of the patch) and global part (of the graph) to generate ground-truth self-supervised signals and learn node representation within graph data. Another work about self-supervised learning for graphs, InfoGraph~\cite{sun2019infograph}, represents each aspect of data by minimizing the distance between the substructure representations with node, edge, triangle scales, etc. Contrastive learning is also performed on the representation learning for the different graphs and node~\cite{hassani2020contrastive} views. There is also a work proposing a pre-training framework to learn the graph representation based on the discrimination of data and contrastive learning~\cite{qiu2020gcc}.

\subsection{\textbf{\chenc{Self-supervised Learning in Sequential Recommendation}}}

Recently, researchers have also integrated self-supervised learning~\cite{hjelm2018learning} into the sequential recommendation~\cite{ma2020disentangled, xin2020self}. In the field of recommendation, more similar to natural language processing, Bert4Rec~\cite{sun2019bert4rec} predicts the random masked items in the sequence by dropout strategy to achieve better generalization. $S^3$-Rec~\cite{zhou2020s3} exploits pre-training to generate self-supervised signals and enhance the representation learning by extracting the ground-truth correlations among attributes, items, sequences, etc. There is also a work about data augmentation, which proposes three strategies to generate self-supervised signals for user pattern extraction and representation learning based on the history item sequences~\cite{xie2020contrastive}. Based on the sequence-to-sequence training strategy, another work performs self-supervised learning and disentanglement on the user intention with history item sequence~\cite{ma2020disentangled}. Besides, there are also some works applying self-supervised learning in general recommendation~\cite{yao2020self} and social recommendation~\cite{tang2013social, yu2021self}. 

Though these self-supervised methods have been proved to be effective, they may not be capable for sequential recommendation because the random dropout strategy would generate sparser data and obscure self-supervision signals for sequential learning. Hence $S^2$-DHCN~\cite{xia2021self} conducts contrastive learning between representations of different hyper-graphs without random dropout, but its fixed ground truths restrict the improvements. CSLR~\cite{zheng2022disentangling} generates self-supervised signals via long-term and short-term interest but fails to further consider more effective signals. Besides, two works~\cite{yu2021socially, xia2021self} about social recommendation~\cite{tang2013social} and session-based recommendation~\cite{GRU4REC, hidasi2018recurrent, chen2019top,  ma2020temporal} address this problem by the self-supervised co-training framework. 

Different from them, our work takes the first step to reflect sequential recommendations from both user-sequence and item-sequence perspectives based on self-supervised learning.

\section{Problem Formulation}\label{sec:proDef}
We first introduce the symbols and then formulate the problem of user-centric, item-centric, and user\&item-centric sequential recommendations to clarify the difference among them. We use $\mathcal{U}_i$ and $\mathcal{I}_u$ to denote the sets of historical users and items in chronological order for the given item $i$ and given user $u$, respectively. Precisely, supposing $u_{t} \in \mathcal{U}_{i_{t+1}}$ and $i_{t} \in \mathcal{I}_{u_{t+1}}$ are the $t$-th user and item that a given target item $i_{t+1}$ and target user $i_{t+1}$ have interacted with. Then the historical sequences of users and items at time step $t$ can be represented as $\mathcal{U}_{i_{t+1}} = (u_{1}, u_{2}, \ldots, u_{t})$ and $\mathcal{I}_{u_{t+1}} = (i_{1}, i_{2}, \ldots, i_{t})$, respectively. Note that at time step $t$, each user or item is not necessary to interact with $t$ items or users. Thus we pad the former number such as $u_{1}$ or $i_{1}$ with 0 (which can be treated as the index for the padding embeddings in the subsequent embedding layer).

\paragraph{\textbf{Sequential Recommendation with User-centric Dimension}}
Given item sequence $\mathcal{I}_{u_{t+1}} = (i_{1}, i_{2}, \ldots, i_{t})$, the user-centric sequential recommendation aims to better predict the probability
that \textbf{given user} $u_{t+1}$ will click the target item \textit{i.e.}, $i_{t+1}$. 
The user-centric sequential recommendation can be formulated as follows.

    \noindent \textbf{Input}: Item sequence $\mathcal{I}_{u_{t+1}} = (i_{1}, i_{2}, \ldots, i_{t})$ for a \textbf{given user $u_{t+1}$}.
    
    \noindent \textbf{Output}: The recommender which predicts the probability that the \textbf{given user $u_{t+1}$} clicks item $i_{t+1}$.

\paragraph{\textbf{Sequential Recommendation with Item-centric Dimension}}
Given user sequence $\mathcal{U}_{i_{t+1}} = (u_{1}, u_{2}, \ldots, u_{t})$, the item-centric sequential recommendation aims to better predict the probability that the \textbf{given item} \textit{i.e.}, $i_{t+1}$ will be clicked by the target user \textit{i.e.}, $u_{t+1}$. 
The item-centric sequential recommendation can be formulated as follows.

\noindent \textbf{Input}: User sequence $\mathcal{U}_{i_{t+1}} = (u_{1}, u_{2}, \ldots, u_{t})$ for a \textbf{given item $i_{t+1}$}.
    
    \noindent \textbf{Output}: The recommender which estimates the probability that the \textbf{given item $i_{t+1}$} will be clicked by the target user $u_{t+1}$.

\paragraph{\textbf{Sequential Recommendation with User\&Item-centric Dimension}}
Unlike the user-centric or item-centric sequential recommendation either focuses on the item or user sequence by the given user or item, the user\&item-centric sequential recommendation considers them simultaneously.
Given user sequence $\mathcal{U}_{i_{t+1}} = (u_{1}, u_{2}, \ldots, u_{t})$ and item sequence $\mathcal{I}_{u_{t+1}} = (i_{1}, i_{2}, \ldots, i_{t})$, the user\&item-centric sequential recommendation aims to better predict the probability
of interaction between the \textbf{given user} \textit{i.e.}, $u_{t+1}$ and the \textbf{given item} \textit{i.e.}, $i_{t+1}$. 

\noindent \textbf{Input}: User sequence $\mathcal{U}_{i_{t+1}} = (u_{1}, u_{2}, \ldots, u_{t})$ and item sequence $\mathcal{I}_{u_{t+1}} = (i_{1}, i_{2}, \ldots, i_{t})$ for a \textbf{given target item $i_{t+1}$ and user $u_{t+1}$}.
    
    \noindent \textbf{Output}: The recommender which estimates the probability of the interaction between the \textbf{given target user $u_{t+1}$ and item $i_{t+1}$}.

\begin{table}[t!]
	\caption{Notation table of important symbols.}
	\label{tbl:notation}

		\begin{centering}
			\setlength\tabcolsep{4pt}   

			\begin{tabular}{ l|l} \toprule
			\textbf{Notations} & \textbf{Descriptions} \\ \hline
				$n, m$ & number of users, number of items\\
$T$ & maximum length of item sequence \\

				$\mathcal{U}_{i_{t+1}} = (u_{1}, u_{2}, \ldots, u_{t})$ &  sequential user sets of the given item $i_{t+1}$ \\
				$\mathcal{I}_{u_{t+1}} = (i_{1}, i_{2}, \ldots, i_{t})$ &  sequential item sets of the given user $u_{t+1}$ \\
				$u_{t} \in \mathcal{U}_{i_{t+1}}$   & the $t$-th user that a given item $i_{t+1}$ has been clicked  \\
				$i_{t} \in \mathcal{I}_{u_{t+1}}$ & the $t$-th item that a given user $u_{t+1}$ has clicked \\
				\hline
				$\mathcal{R}$ & training set \\
			    $\mathcal{R}^{te}$ & test set \\
		
				$y_{u_{t+1}, i_{t+1}} \in \{0,1\}$ &  
				$			\begin{cases}
					1&  \text{if user } u_{t+1} \text{ clicks item } i_{t+1} \\
					0&  \text{if user } u_{t+1} \text{ does not click item } i_{t+1}
				\end{cases}$ \\

				\hline
				$D$ & number of latent dimensionality \\

	$\textbf{M}^u \in \mathbb{R}^{n  \times D}$  & user embedding matrix\\
	$\textbf{M}^i \in \mathbb{R}^{m  \times D}$  & item embedding matrix\\
	$\mathbf{E}^{u}$, $\mathbf{E}^{i} \in \mathbb{R}^{T \times D}$ & the user and item embeddings for the model inputs \\

$\boldsymbol{h}^u_{i_{t+1}} \in \mathbb{R}^{1  \times D}$ & the hidden states for the user sequence of target item $i_{t+1}$  \\
 $\boldsymbol{h}^i_{u_{t+1}} \in \mathbb{R}^{1  \times D}$ & the hidden states for the item sequence of target user $u_{t+1}$ \\
 ${\textbf{U}}^i_{u_{t+1}} \in \mathbb{R}^{1  \times D}$ & user representation under the item sequence of target user $u_{t+1}$\\
 ${\textbf{V}}^u_{i_{t+1}} \in \mathbb{R}^{1  \times D}$ & item representation under the user sequence of target item $i_{t+1}$\\
\hline
$\lambda_e $ & L2 regularization parameter for the representation contrastive learning \\
$\lambda_p $ & L2 regularization parameter for the interest contrastive learning \\
				\bottomrule
			\end{tabular}
			
		\end{centering}
	
\end{table}
\section{Methodology}\label{sec:method}

\subsection{Overview of DCN}

\begin{figure}[t!]
		\centering
		\includegraphics[width=0.95\linewidth]{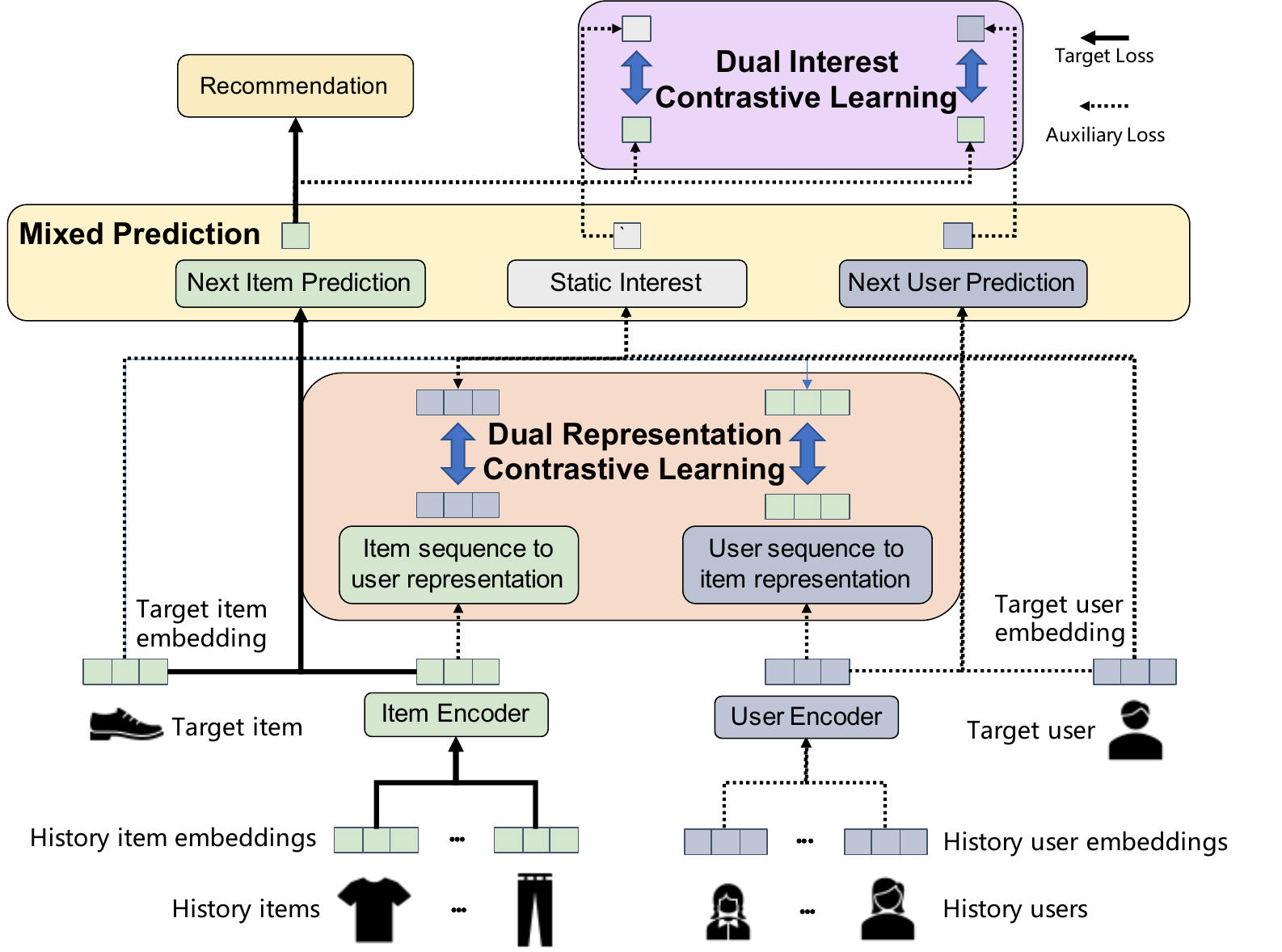}
	\caption{Illustration of our DCN. (1) Item embeddings and user embeddings are looked up based on the item sequence and user sequence before being encoded by the item encoder and user encoder to capture the sequential transition patterns; (2) \textbf{Dual Representation Contrastive Learning} self-supervises the output of item encoder with the target user embedding and that of user encoder with the target item embedding; (3) \textbf{Mixed Prediction Layer} predicts the probabilities of next item prediction, static interest, and next user prediction; (4) \textbf{Dual Interest Contrastive Learning} self-supervises the dynamic interest of next item prediction with the target user-item interaction prediction and the next user prediction.}	\label{fig:architecture}
\end{figure} 

Our DCN model is illustrated in Figure~\ref{fig:architecture}, enhancing the sequential recommendation with contrastive learning from the representation aspect and interest aspect based on the novel auxiliary user sequence. The proposed model mainly consists of the following four parts:

\begin{itemize}[leftmargin=*]

\item \textbf{Sequential Encoding Layer}.
We build a user encoder and item encoder to capture the user-to-user transition pattern and item-to-item transition pattern for the given target item and user.  

\item \textbf{Dual Representation Contrastive Layer}. 
Existing sequential recommendation only focuses on the user-centric view, which results in the overfitting of item representation in the item sequence. Thus we treat the outputs of the user encoder and item encoder as the representations for the target item and target user, respectively, and propose dual representation contrastive learning to self-supervise them with their embeddings to refine the representation learning. 

 \item \textbf{Mixed Prediction Layer}. 
Here we tend to achieve the next item and user prediction tasks with additional traditional collaborative filtering. For the dynamic interest, we predict the next user (next item) by feeding the target user (item) embedding with the output of the user encoder (item encoder) into the prediction layer. For the static interest, we predict the interaction between the target user and the target item by feeding their embeddings into the prediction layer.
 \item \textbf{Dual Interest Contrastive Layer}. 
 We tend to incorporate next-user prediction and collaborative filtering tasks for more supervised signals.
Specifically, we propose dual interest contrastive learning to self-supervise the dynamic interest for next-item prediction with the static interest. Besides, we also self-supervise the predicted next user score with the predicted next item score.
\end{itemize}

\chenc{In short, the user and item will interact in three modules. Firstly, in the Dual Representation Contrastive Layer, the item sequence will interact with the target user while the user sequence will interact with the target item. Secondly, in the Mixed Prediction Layer, the target user will interact with the target item. Finally, in the Dual Interest Contrastive Layer, the predicted next item score will interact with the predicted next user score.}

\subsection{Sequential Encoding Layer}

\subsubsection{\textbf{Embedding Layer}}

We create a user embedding matrix $\textbf{M}^u \in \mathbb{R}^{n  \times D}$ and an item embedding matrix $\textbf{M}^i \in \mathbb{R}^{ m \times D}$ where $n$ and $m$ denote the numbers of users and items, respectively. Here $D$ denotes the latent dimensionality. \chenc{Note that though we define all the hidden states and embeddings with the same dimension size $D$, they are unnecessary to be with the same size. Here we do so to simplify the illustration following existing works~\cite{DIN, SASRec, SURGE}. Besides, experiments have shown that keeping them the same is beneficial to fair comparison with baselines.} Given user sequence $\mathcal{U}_{i_{t+1}} = (u_{1}, u_{2}, \ldots, u_{t})$ and item sequence $\mathcal{I}_{u_{t+1}} = (i_{1}, i_{2}, \ldots, i_{t})$, we lookup the input embeddings as:
\begin{equation}\label{eqn:emb}
\begin{aligned}
\mathbf{E}^{u} &= [\textbf{M}^u_{u_{1}} , \textbf{M}^u_{u_{2}} , \ldots, \textbf{M}^u_{u_{t}} ], \\
\mathbf{E}^{i} &= [\textbf{M}^i_{i_{1}} , \textbf{M}^i_{i_{2}} , \ldots, \textbf{M}^i_{i_{t}} ],\\
\end{aligned}
\end{equation}
where $\mathbf{E}^{u}$ and $\mathbf{E}^{i} \in \mathbb{R}^{T \times D}$ denote the user and item embeddings for the model inputs, respectively. Note that if the sequence length is less than $t$, we will pad them with zero embedding~\cite{SASRec}.

\subsubsection{\textbf{User Encoder and Item Encoder}}
The traditional sequential recommendation often exploits an item encoder solely to capture the item-to-item transition pattern of a certain item sequence, only considering the user-centric dimension. Here we additionally exploit the user encoder to further capture the user-to-user transition pattern, further considering the item-centric dimension and achieving mutual harmony.
As we have obtained $\mathbf{E}^{u}$ and $\mathbf{E}^{i}$ from the above embedding layers, we then exploit two single sequential models as user encoder and item encoder to learn the  sequential patterns for user sequence and item sequence, respectively, as follows,
\begin{equation}\label{eq:localA} 
\begin{aligned}
	\boldsymbol{h}^u_{i_{t+1}}= \textbf{Encoder}_u(\mathbf{E}^u),\\
	\boldsymbol{h}^i_{u_{t+1}}= \textbf{Encoder}_i(\mathbf{E}^{i}),\\
\end{aligned}
\end{equation}
where $\boldsymbol{h}^u_{i_{t+1}}$ and $\boldsymbol{h}^i_{u_{t+1}}$ are the hidden states for the user sequence of target item $i_{t+1}$ and the item sequence of target user $u_{t+1}$, respectively. Here $\textbf{Encoder}_u$ and $\textbf{Encoder}_i$ are the sequential encoders for user sequence and item sequence, respectively, which can be GRU4REC~\cite{GRU4REC}, SASRec~\cite{SASRec} and  SLi-Rec~\cite{SLIREC} etc. 
Here take the \textbf{GRU4REC} backbone as an example, which can be defined as follows if applying it as a user encoder,
\begin{equation}\label{eq:gru_u}
\begin{aligned}
&\mathbf{Z}^u_t =\sigma\left(\mathbf{W}^u_{z}\left[\mathbf{E}^{u}_t, \mathbf{H}^u_{t-1}\right]\right), \\
&\mathbf{R}^u_{t} =\sigma\left(\mathbf{W}^u_{r}\left[\mathbf{E}^{u}_t, \mathbf{H}^u_{t-1}\right]\right), \\
&\tilde{\mathbf{H}}^u_t =\tanh \left(\mathbf{W}^u_{\widetilde{h}}\left[\mathbf{E}^{u}_t, \mathbf{R}^u_{t} \odot \mathbf{H}^u_{t-1}\right]\right),  \\
&\mathbf{H}^u_t =\left(1-\mathbf{Z}^u_t\right) \odot \mathbf{H}^u_{t-1}+\mathbf{Z}^u_t \odot \tilde{\mathbf{H}}^u_t, 
\end{aligned}
\end{equation}
when \textbf{GRU4REC} backbone is used as an item encoder, we can have:
\begin{equation}\label{eq:gru_i}
\begin{aligned}
&\mathbf{Z}^i_t =\sigma\left(\mathbf{W}^i_{z}\left[\mathbf{E}^{i}_t, \mathbf{H}^i_{t-1}\right]\right), \\
&\mathbf{R}^i_{t} =\sigma\left(\mathbf{W}^i_{r}\left[\mathbf{E}^{i}_t, \mathbf{H}^i_{t-1}\right]\right), \\
&\tilde{\mathbf{H}}^i_t =\tanh \left(\mathbf{W}^i_{\widetilde{h}}\left[\mathbf{E}^{i}_t, \mathbf{R}^i_{t} \odot \mathbf{H}^i_{t-1}\right]\right),  \\
&\mathbf{H}^i_t =\left(1-\mathbf{Z}^i_t\right) \odot \mathbf{H}^i_{t-1}+\mathbf{Z}^i_t \odot \tilde{\mathbf{H}}^i_t, 
\end{aligned}
\end{equation}
where $\mathbf{W}^u_{z}$, $\mathbf{W}^u_{r}$ and $\mathbf{W}^u_{\widetilde{h}} \in \mathbb{R}^{D \times 2D}$ and $\mathbf{W}^i_{z}$, $\mathbf{W}^i_{r}$ and $\mathbf{W}^i_{\widetilde{h}} \in \mathbb{R}^{D \times 2D}$ are parameters to be learned. Here $\odot$ is the Hadamard product.

Based on it, we obtain the hidden states $\mathbf{H}^u$ and $\mathbf{H}^i \in \mathbb{R}^{T \times D}$ which capture user sequential pattern and item sequential pattern, respectively. Given user sequence $\mathcal{U}_{i_{t+1}} = (u_{1}, u_{2}, \ldots, u_{t})$ and item sequence $\mathcal{I}_{u_{t+1}} = (i_{1}, i_{2}, \ldots, i_{t})$ for a given target item $i_{t+1}$ and target user $i_{t+1}$, we represent them as below.
\begin{equation}\label{eq:rep} 
	\boldsymbol{h}^u_{i_{t+1}}= \mathbf{H}^u_t,
	\boldsymbol{h}^i_{u_{t+1}}= \mathbf{H}^i_t
\end{equation}
where $\boldsymbol{h}^u_{i_{t+1}}$ and $\boldsymbol{h}^i_{u_{t+1}}$ are the representations for the user sequence $\mathcal{U}_{i_{t+1}} = (u_{1}, u_{2}, \ldots, u_{t})$ and item sequence $\mathcal{I}_{u_{t+1}} = (i_{1}, i_{2}, \ldots, i_{t})$ for a given target item $i_{t+1}$ and target user $i_{t+1}$, respectively, as Eq.\eqref{eq:localA}.

\subsection{Dual Representation Contrastive Layer}
As traditional sequential recommendation often relays on item sequence and overfits in a user-centric view, we tend to refine the representation learning with embeddings from an item-centric view. Specifically, in this section, we perform a contrastive learning layer from the representation aspect as the middle part of  Figure~\ref{fig:architecture}. We design such a contrastive learning approach based on the intuition that an item sequence can be referred to as a user and a user sequence can be referred to as an item.
\subsubsection{\textbf{Item Sequence to User Representation Transformation}}
Each item sequence of a given user can represent his/her specific preference or intention. \chenc{Directly representing user by the item sequential embeddings may confuse the learning process as they will be leveraged to extract item sequential patterns subsequently}. Thus we transform item sequential embeddings into a user representation space to represent the given user.
Specifically, after obtaining the sequential embeddings $\boldsymbol{h}^i_{u_{t+1}}$ under item sequence $\mathcal{I}_{u_{t+1}} = (i_{1}, i_{2}, \ldots, i_{t})$, we project them into the target user $u_{t+1}$ space by the following equation.
\begin{equation}\label{eq:uti_space}
\begin{aligned}
 {\textbf{U}}^i_{u_{t+1}} = {\boldsymbol{h}}^i_{u_{t+1}} \textbf{W}^{i} + \boldsymbol{b}^{i},
\end{aligned}
\end{equation}
where $\textbf{W}^{i} \in \mathbb{R}^{D \times D}$ and $\boldsymbol{b}^{i} \in \mathbb{R}^{D}$ are parameters to be learned. We then treat ${\textbf{U}}^i_{u_{t+1}}$ as the representation for target user $u_{t+1}$. 
\subsubsection{\textbf{User Sequence to Item Representation Transformation}}
Likewise, each user sequence of a given item can represent its specific preferred trend. Thus we transform user sequential embeddings into an item representation space to represent the given item.
Specifically, after obtaining the sequential embeddings $\boldsymbol{h}^u_{i_{t+1}}$ under user sequence $\mathcal{U}_{i_{t+1}} = (u_{1}, u_{2}, \ldots, u_{t})$, we project them into the target item $i_{t+1}$ space by the following equation.
\begin{equation}\label{eq:itu_space}
\begin{aligned}
 {\textbf{V}}^u_{i_{t+1}} = {\boldsymbol{h}}^u_{i_{t+1}} \textbf{W}^{u} + \boldsymbol{b}^{u},
\end{aligned}
\end{equation}
where $\textbf{W}^{u} \in \mathbb{R}^{D \times D}$ and $\boldsymbol{b}^{u} \in \mathbb{R}^{D}$ are parameters to be learned. We then treat ${\textbf{V}}^u_{i_{t+1}}$ as the representation for target item $i_{t+1}$. 
\subsubsection{\textbf{Dual Representation Contrastive Learning}}
This contrastive learning approach tends to refine the target user and item embeddings via the encoded sequential representations for the target user and target item, respectively.
After obtaining the representations ${\textbf{V}}^i_{u_{t+1}}$ and ${\textbf{U}}^u_{i_{t+1}}$ for the target item and user, respectively, we further perform contrastive learning on them with the target item embedding $\textbf{M}^i_{i_{t+1}}$ and target user embedding $\textbf{M}^u_{u_{t+1}}$, as:
\begin{equation}\label{eq:dr}
\mathcal{L}^{e}= \lambda_e \left( \left({\textbf{U}}^u_{i_{t+1}} - \textbf{M}^i_{i_{t+1}}\right)^2 + \left({\textbf{V}}^i_{u_{t+1}} - \textbf{M}^u_{u_{t+1}}\right)^2\right), 
\end{equation}
where $\lambda_e $ is the L2 regularization parameter for contrastive representation learning.
After the dual representation contrastive learning layer, we obtain refined representations for the item and user sequences.

\subsection{Mixed Prediction Layer}
As the problem formulation section illustrates, we can achieve the next item and user prediction from a user-centric and item-centric perspective, respectively, based on the history item and user sequence. Besides, in this section, we also predict the machining probability. These three predictions are achieved via different MLP towers and will be linked together to generate ground-truth self-supervised signals in the subsequent section.
With the outputs from the user encoder and item encoder, we concatenate them together and feed them into the MLP-based prediction layer~\cite{DIN, DIEN}, which can be formulated as follows,
\begin{equation}
	P\left(u_{t + 1}|u_{1}, u_{2}, \ldots, u_{t}\right)=\textbf{MLP}\left(\boldsymbol{h}^u_{i_{t+1}} \|\mathbf{M}^u_{u_{t+1}}\right),
\end{equation}
 \begin{equation}
	P\left(i_{t + 1}|i_{1}, i_{2}, \ldots, i_{t}\right)=\textbf{MLP}\left(\boldsymbol{h}^i_{u_{t+1}} \|\mathbf{M}^i_{i_{t+1}}\right),
\end{equation}
\begin{equation}
	P\left(u_{t + 1},i_{t + 1}\right)=\textbf{MLP}\left(\boldsymbol{h}^u_{i_{t+1}} \|\boldsymbol{h}^i_{u_{t+1}}\right),
\end{equation}
where $P\left(u_{t + 1}|u_{1}, u_{2}, \ldots, u_{t}\right)$, $P\left(i_{t + 1}|i_{1}, i_{2}, \ldots, i_{t}\right)$ and $P\left(u_{t + 1},i_{t + 1}\right)$ are the predicted scores (probabilities) of next user prediction, next item prediction and static interest. These prediction probabilities will be contrasted to generate self-supervised signals in the subsequent section.

\subsection{Dual Interest Contrastive Layer}
Extracting item sequential pattern only from user-centric view, existing sequential recommendation often aims to capture the dynamic interest of the specific user. However, it fails to capture the dynamic trend of the specific item being consumed. Thus we further perform the next user prediction in the previous section to capture the dynamic trend of the specific item. Apart from the next item and user predictions, we also predict the machining probability. We define these three prediction tasks as different interest modeling, e.g., next item and user predictions as dynamic interest and the machining probability as static interest. Performing contrastive learning on these three tasks can provide more self-supervised signals to the training process and address the problem of unreliable supervised signals from sparse data for the traditional sequential recommendation. In this section, we tend to perform a contrastive layer at the interest (prediction probability) aspect.

After the mixed prediction layer, we obtain the predicted scores of next user prediction, next item prediction, and static interest. We then tend to mix them further together for the same optimization objective.
\subsubsection{\textbf{Dual Interest Contrastive Learning}}
As the representation contrastive learning refines the item embeddings from a user-centric view with user embeddings from an item-centric view based on unsupervised contrastive learning, the traditional sequential recommendation is still short of supervised signals.
Thus, apart from the contrastive representation learning, we further propose dual interest contrastive learning on the dynamic interest for next item prediction with the static interest as follows.
\begin{equation}\label{eq:di}
\begin{aligned}
\mathcal{L}^{p}= &\lambda_p  ( \left(P\left(i_{t + 1}|i_{1}, i_{2}, \ldots, i_{t}\right) - P\left(u_{t + 1},i_{t + 1}\right)\right)^2 + \\
&\left(P\left(i_{t + 1}|i_{1}, i_{2}, \ldots, i_{t}\right) - P\left(u_{t + 1}|u_{1}, u_{2}, \ldots, u_{t}\right)\right)^2)  
\end{aligned}
\end{equation}
where $\lambda_p$ is the L2 regularization parameter for the interest contrastive learning. Here next user score is also self-supervised.
With contrastive learning on the prediction probabilities, there will be more self-supervised signals for the backpropagation of the gradient.
\subsubsection{\textbf{Model Optimization}}
Following the popular next item prediction, we then utilize the widely-used \textit{LogLoss} function~\cite{DIN, DIEN} as the main loss function, formulated as follows, 
\begin{equation}\label{eq:loss}
\begin{aligned}
\mathcal{L}^i=&-\frac{1}{|\mathcal{R}|} \sum_{(u_{t+1}, i_{t+1}) \in \mathcal{R}}(y_{u_{t+1}, i_{t+1}} \log P\left(i_{t + 1}|i_{1}, i_{2}, \ldots, i_{t}\right) +\\
&\left(1-y_{u_{t+1}, i_{t+1}}\right) \log \left(1-P\left(i_{t + 1}|i_{1}, i_{2}, \ldots, i_{t}\right)\right)),
\end{aligned}
\end{equation}
where $\mathcal{R}$ is the training set. 
Here $y_{u_{t+1}, i_{t+1}}=1$  and $y_{u_{t+1}, i_{t+1}}=0$  denote the positive/negative training samples, respectively.

To jointly optimize the next item prediction with the next user prediction and static interest, our final loss function combine $\mathcal{L}^i$ with $\mathcal{L}^e$ and $\mathcal{L}^p$ in \chenc{Eqn.\eqref{eq:dr} and Eqn.\eqref{eq:di}}.
It can be formulated as follows,
\begin{equation}\label{eq:loss}
\mathcal{L}= \mathcal{L}^i + \mathcal{L}^e + \mathcal{L}^{p} + \lambda\|\Theta\|_{2},
\end{equation}
where $\Theta$ denotes learnable parameters with hyper-parameter $\lambda$ for the regularization penalty. Based on this joint loss, we can refine the representation learning, generate more reliable self-supervised signals and prevent overfitting for the traditional sequential recommendation.

\noindent \textbf{Discussion}. Our early attempt has discovered that predicting the next user based on the historical user sequence of a given target item is poorer than the next item prediction. This may be because the sequential behavior data is generated by user-centric next item prediction, and the traditional sequential recommendation is often evaluated on item ranking instead of user ranking. 
Therefore, mere user sequence is not suitable to achieve traditional sequential recommendation tasks, which may be the reason why existing works do not attempt the next user prediction. But indeed, item-centric recommendation with next user prediction can be an effective supplement for the traditional user-centric recommendation with next item recommendation.

\section{Experiments}\label{sec:exp}
In the experimental section, we tend to conduct extensive experiments with three public datasets and one industrial datasets. Specifically, the subsequent four research questions (RQs) are investigated.
\begin{itemize}[leftmargin=*]
	\item \textbf{RQ1:} When achieving mutual harmony, does the proposed method outperform the state-of-the-art sequential recommendation models on both user-centric and item-centric evaluation metrics?
		\item \textbf{RQ2:}  What is the contribution of different proposed components to the overall performance? 
		\item \textbf{RQ3:} Are the proposed components model-agnostic? What is their generalization ability of them on different sequential backbones?
				\item \textbf{RQ4:}  What is the optimal value of the regularization parameter for the contrastive learning loss on user-centric and item-centric evaluation metrics?
\end{itemize}
\subsection{Experimental Setup}
\subsubsection{\chenc{\textbf{Datasets}}}
\begin{table}[t!]
\small
\caption{Data statistics for four datasets \chenc{after filtering out users and items with less than 10 interactions}.}\label{tbl:data}

\begin{tabular}{c|c|c|c|c|c|c}
\hline
\multirow{2}{*}{\textbf{Dataset}} & \multirow{2}{*}{\textbf{\#Users}} & \multirow{2}{*}{\textbf{\#Items}} & \multirow{2}{*}{\textbf{\#Records}} & \multirow{2}{*}{\textbf{Density}} & \multirow{2}{*}{\textbf{\begin{tabular}[c]{@{}c@{}}Average records \\      per user\end{tabular}}} & \multirow{2}{*}{\textbf{\begin{tabular}[c]{@{}c@{}}Average records\\       per item\end{tabular}}} \\
                                  &                                   &                                   &                                     &                                   &                                                                                                 &                                                                                                 \\ \hline
\textbf{Amazon Toys}              & 6,919                             & 28,695                            & 120,334                             & 0.06\%                            & 17.39                                                                                           & 4.10                                                                                            \\ \hline
\textbf{Amazon Games}             & 6,955                             & 15,316                            & 133,121                             & 0.12\%                            & 19.14                                                                                           & 8.69                                                                                            \\ \hline
\textbf{Taobao}                   & 37,293                            & 64,996                            & 1,505,878                           & 0.06\%                            & 40.38                                                                                           & 23.17                                                                                           \\ \hline
\textbf{Micro-video}              & 37,497                            & 128,758                           & 6,413,396                           & 0.13\%                            & 171.04                                                                                          & 49.81                      \\ \hline                                                                    
\end{tabular}
\end{table}

We conduct evaluations of recommendation performance on four benchmark datasets. 
The statistics of the filtered datasets with a 10-core setting are shown in Table~\ref{tbl:data}, where Avg. records per user and per item are the average length of item sequence and user sequence for user-centric and item-centric sequential modeling, respectively.
\begin{itemize}[leftmargin=*]
\item \textbf{Amazon Toys}\footnote{https://www.amazon.com}. This highly sparse dataset is collected from the Toys domain of the largest e-commerce platform around the world, where user ratings are ranged from 1 to 5. We split the behaviors before the last year as the training set, the first half of the last year as the validation set, and the second half of the last year as the test set. 
\item \textbf{Amazon Games}. This is also a highly sparse dataset and is collected from the Games domain of the largest e-commerce platform around the world. We split the training, validation, and test set following the Toys domain.

\item \textbf{Taobao}\footnote{https://tianchi.aliyun.com/dataset/dataDetail?dataId=649}. This dataset is public and provided by an online e-commerce platform. We filter the instances before November 25 and after December 3, 2017. We split the instances of first seven days for training, the 8-th day for validation, and the 9-th day for testing, following the existing work~\cite{SURGE}.

\item \textbf{Micro Video}\chenc{\footnote{\chenc{\url{ http://bit.ly/3BR8rqg}}}}. This dataset is sampled from one of the largest micro-video apps \chenc{Kuaishou} in China, where users can share their videos. The videos viewed, liked, followed, and skipped are recorded in the dataset. We downsample the logs from September 11 to September 22, 2021, and select the click data before filtering out users and videos which are interacted less than ten times via the 10-core setting. We split the behaviors before 12 pm on the last day and after 12 pm on the last day, respectively, as the validation set and test set. Other behaviors are used as the training set.
\end{itemize}

Notice that we process all datasets following CTR~\cite{DIN, SURGE} setting , which is different from the processing of datasets in SASRec~\cite{SASRec}.

\subsubsection{\chenc{\textbf{Baseline Models}}}
To demonstrate the effectiveness of our model, we compare it with \chenc{three categories of representative and competitive baselines: non-sequential models that capture user's static preference, sequential models that further capture sequential transition patterns, and contrastive sequential models further exploit contrastive learning to promote the sequential recommendation}. 

\noindent \textbf{Non-sequential Models}
\begin{itemize}[leftmargin=*]
    \item \textbf{NCF}~\cite{NCF}: NCF is the state-of-the-art non-sequential recommendation method that incorporates matrix factorization with multi-layer perceptron (MLP) to predict user-item machining probability by replacing the inner product on the user and item embeddings with a neural network.
        \item \chenc{\textbf{LightGCN}~\cite{lightgcn}: LightGCN leverages and simplifies  graph convolution network to capture the higher-order relation of recommendation.}
        
    \item \textbf{DIN}~\cite{DIN}: DIN represents the user by aggregating his/her weighted historical item embeddings based on attention weights which are calculated by feeding the target item embedding with each history item embedding into the MLP.
\end{itemize}

\noindent \textbf{Sequential Models}
\begin{itemize}[leftmargin=*]

    \item \textbf{Caser} ~\cite{Caser}: Caser performs horizontal and vertical convolutions on user's item sequential representation to extract the union-level patterns and point-level sequential patterns, respectively.
    \item \textbf{GRU4REC}~\cite{GRU4REC}: GRU4REC exploits GRU (Gated Recurrent Unit) ~\cite{GRU} to predict the next interacted items in a given session by the final state of the unit output. GRU is a gated unit to address the gradient vanishing problem of RNN (Recurrent Neural Network)~\cite{mikolov2010recurrent}.
    
    \item \textbf{DIEN}~\cite{DIEN}: Improved based on DIN, DIEN proposes two GRU layers for interest extraction and evolution, respectively, to extract dynamic interests from history item sequence and interest hidden state that is relative to the target item. 
    
\item \textbf{SASRec}~\cite{SASRec}: SASRec encodes the item sequence via the hierarchical self-attention network, which is supposed to capture the long-term relationships among all history items.
\item \textbf{SLi-Rec}~\cite{SLIREC}: SLi-Rec exploits DIN as an attention network to model the long-term interest, which is static or evolving slowly, and time-aware LSTM to model the short-term interest, which is dynamic or evolving frequently.
\item \textbf{SURGE}~\cite{SURGE}: SURGE firstly leverages metric learning to construct an attention graph based on the historical item sequence.
Then it adopts convolution and pooling operations to aggregate the constructed item graph.

\end{itemize}
\chenc{
\noindent \textbf{Contrastive Sequential Models}}
\begin{itemize}[leftmargin=*]
\item \chenc{\textbf{BERT4Rec}~\cite{sun2019bert4rec}: BERT4Rec leverages a dropout strategy from NLP that randomly masks the items while training. It is a stack of bi-directional self-attention mechanism for sequential recommendation task.}

\item \chenc{\textbf{$S^3$-Rec}~\cite{zhou2020s3}: $S^3$-Rec also leverages dropout strategy for training, which further exploits variant in masked item prediction.}
\item \chenc{\textbf{DuoRec}~\cite{duorec}: DuoRec proposes a contrastive method to regularize the sequence representations and a dropout augmentation method to better preserve the semantic information. Besides, it develops a sampling strategy to choose hard positive samples based on item sequences with the same target item.}
\end{itemize}

\subsubsection{\chenc{\textbf{User-centric and Item-centric Evaluation Metrics}}}

We use AUC and GAUC (two widely-used accuracy metrics~\cite{gunawardana_evaluating_2015}) and MRR and NDCG@10 (two widely-used ranking metrics~\cite{SURGE}) for both user-centric and item-centric evaluation. Note that we rank the top items in the user-centric evaluation while ranking the top users in the item-centric evaluation. The details of the adopted evaluation metrics are as below.
\begin{itemize}[leftmargin=*]
     \item \textbf{AUC} calculates the probability that the prediction scores of positive target items are ranking higher than those of negative items, evaluating the model's accuracy of classification performance~\cite{gunawardana_evaluating_2015}.
     \item \textbf{GAUC} further calculates each user's AUC by weighted average, where the weight is the number of his/her clicked items~\cite{gunawardana_evaluating_2015}. It evaluates the model performance in a more bias-aware and fine-grained manner when the recommender systems indeed tend to rank the top items for each certain user.
     \item \textbf{MRR} (short for Mean Reciprocal Rank) averages the first correct item's inverse ranking.
     
     \item \textbf{NDCG@K} thinks highly of those items at topper positions in the recommended top $K$ items, where the test items rank higher will result in better evaluating performance. In our experiments, $K$ is set to 10, a popular setting in related work~\cite{SASRec}\chenc{, which means we will sample nine negative items while testing each positive item}.
 \end{itemize}
\subsubsection{\chenc{\textbf{Hyper-parameter Settings}}}
We follow an existing work about sequential recommendation~\cite{SURGE}, all models are optimized by Adam with a learning rate of $0.001$~\cite{Adam}, whose parameters are initialized by Xavier initialization~\cite{xavier}. \chenc{Here we have also initialized the learning rate with different value, which makes no significant difference to the final results.} We search the regularization parameters for penalty on model embedding within $[10^{-7}$, $10^{-5}$, $10^{-3}]$. Besides, we also search for the best regularization parameters for contrastive learning as Section~\ref{sec:regularization_para}, where the selected best parameters are as shown in Table~\ref{tbl:best_para}.
As for other parameters, \chenc{all} datasets are fed with batch size 200. The embedding sizes are fixed as 32 for all models. The prediction layers of all models are two-layer MLPs with layer sizes 100 and 64. The maximum sequence length of the Taobao is 50, while that of the Amazon Toys is 20. 

All the models are implemented based on Python with a TensorFlow~\footnote{ https://www.tensorflow.org} framework of Microsoft~\footnote{https://github.com/microsoft/recommenders} for recommendation tasks and tuned with the best parameters.
\begin{table}[t!]
\caption{Best regularization parameters for contrastive learning.}\label{tbl:best_para}
\begin{tabular}{c|c|c|c|c}
\hline
\textbf{Model}   & \textbf{Amazon Toys} & \textbf{Amazon Games} & \textbf{Taobao} & \textbf{Micro-video} \\ \hline
\textbf{GRU4REC} & $10^{-5}$              & $10^{-6}$               & $10^{-3}$         & $10^{-4}$              \\ \hline
\textbf{SASRec}  & $10^{-5}$              & $10^{-6}$               & $10^{-5}$         & $10^{-4}$              \\ \hline
\textbf{SLi-Rec} & $10^{-5}$              & $10^{-5}$               & $10^{-3}$         & $10^{-3}$              \\ \hline
\end{tabular}
\end{table}

\begin{table}[t!]
\caption{Item recommendation performance comparisons on Amazon Toys, Amazon Games, Taobao, and Micro-video \chenc{(bold means $p$-value < 0.05, bold$^{*}$ means $p$-value < 0.01, and bold$^{**}$ means $p$-value < 0.001). }}\label{tbl:overall_item}
\setlength\tabcolsep{3pt}
\chenc{
\begin{tabular}{ccccccccc}
\hline
\textbf{Dataset} & \multicolumn{4}{c}{\textbf{Amazon Toys}} & \multicolumn{4}{c}{\textbf{Amazon Games}} \\ \hline
\textbf{Metric} & \textbf{AUC} & \textbf{MRR} & \textbf{NDCG} & \textbf{WAUC} & \textbf{AUC} & \textbf{MRR} & \textbf{NDCG} & \textbf{WAUC} \\ \hline
\textbf{NCF} & 0.6162 & 0.3791 & 0.5240 & 0.6192 & 0.7258 & 0.5049 & 0.6229 & 0.7343 \\ \hline
\textbf{LightGCN} & 0.6249 & 0.3827 & 0.5271 & 0.6278 & 0.6580 & 0.4185 & 0.5553 & 0.6610 \\ \hline
\textbf{DIN} & 0.6647 & 0.4694 & 0.5942 & 0.6914 & 0.7987 & {\ul 0.6447} & {\ul 0.7306} & {\ul 0.8284} \\ \hline
\textbf{Caser} & 0.6676 & 0.4708 & 0.5958 & 0.6996 & 0.7908 & 0.6210 & 0.7125 & 0.8138 \\ \hline
\textbf{GRU4REC} & 0.7245 & 0.4835 & 0.6059 & 0.7151 & 0.8130 & 0.6437 & 0.7299 & 0.8270 \\ \hline
\textbf{DIEN} & 0.6624 & 0.4665 & 0.5921 & 0.6902 & 0.7866 & 0.6311 & 0.7203 & 0.8226 \\ \hline
\textbf{SASRec} & 0.6800 & 0.4789 & 0.6020 & 0.7042 & 0.7707 & 0.5824 & 0.6830 & 0.7926 \\ \hline
\textbf{SLi-Rec} & 0.7234 & 0.5054 & 0.6228 & 0.7304 & 0.8067 & 0.6408 & 0.7273 & 0.8193 \\ \hline
\textbf{SURGE} & 0.7262 & {\ul 0.5079} & 0.6240 & 0.7313 & 0.8072 & 0.6412 & 0.7275 & 0.8195 \\ \hline
\textbf{BERT4Rec} & 0.6806 & 0.4763 & 0.6002 & 0.7073 & 0.7635 & 0.6224 & 0.7135 & 0.8127 \\ \hline
\textbf{$S^3$-Rec} & 0.6913 & 0.4861 & 0.6114 & 0.7181 & 0.7742 & 0.6331 & 0.7239 & 0.8219 \\ \hline
\textbf{DuoRec} & {\ul 0.7342} & 0.5072 & {\ul 0.6252} & {\ul 0.7324} & {\ul 0.8151} & 0.6439 & 0.7301 & 0.8221 \\ \hline
\textbf{DCN} & \textbf{0.7471$^{**}$} & \textbf{0.5110$^{*}$} & \textbf{0.6274$^{*}$} & \textbf{0.7371$^{*}$} & \textbf{0.8235$^{**}$} & \textbf{0.6600$^{**}$} & \textbf{0.7422$^{**}$} & \textbf{0.8351$^{*}$} \\ \hline
\textbf{Dataset} & \multicolumn{4}{c}{\textbf{Taobao}} & \multicolumn{4}{c}{\textbf{Micro-video}} \\ \hline
\textbf{Metric} & \textbf{AUC} & \textbf{MRR} & \textbf{NDCG} & \textbf{WAUC} & \textbf{AUC} & \textbf{MRR} & \textbf{NDCG} & \textbf{WAUC} \\ \hline
\textbf{NCF} & 0.7409 & 0.5509 & 0.6584 & 0.7684 & 0.5244 & 0.3080 & 0.4672 & 0.5327 \\ \hline
\textbf{LightGCN} & 0.5736 & 0.3358 & 0.4898 & 0.5767 & 0.7374 & 0.5287 & 0.6448 & 0.7538 \\ \hline
\textbf{DIN} & 0.7888 & 0.6767 & 0.7552 & 0.8491 & 0.7289 & {\ul 0.5524} & {\ul 0.6598} & 0.7546 \\ \hline
\textbf{Caser} & 0.8165 & 0.6670 & 0.7476 & 0.8415 & 0.7223 & 0.4903 & 0.6119 & 0.7250 \\ \hline
\textbf{GRU4REC} & 0.8468 & 0.6940 & 0.7681 & 0.8544 & 0.6928 & 0.4683 & 0.5941 & 0.6959 \\ \hline
\textbf{DIEN} & 0.8022 & 0.6812 & 0.7586 & 0.8510 & 0.6994 & 0.5135 & 0.6295 & 0.7186 \\ \hline
\textbf{SASRec} & 0.8086 & 0.6310 & 0.7204 & 0.8260 & 0.7344 & 0.5263 & 0.6400 & 0.7515 \\ \hline
\textbf{SLi-Rec} & 0.8584 & 0.7077 & 0.7786 & 0.8624 & 0.7276 & 0.5130 & 0.6295 & 0.7335 \\ \hline
\textbf{SURGE} & 0.8592 & 0.7080 & {\ul 0.7811} & {\ul 0.8652} & 0.7308 & 0.5158 & 0.6309 & 0.7345 \\ \hline
\textbf{BERT4Rec} & 0.8108 & 0.6328 & 0.7227 & 0.8280 & 0.7488 & 0.5328 & 0.6545 & 0.7610 \\ \hline
\textbf{$S^3$-Rec} & 0.8194 & 0.6397 & 0.7319 & 0.8359 & {\ul 0.7502} & 0.5362 & 0.6562 & {\ul 0.7633} \\ \hline
\textbf{DuoRec} & {\ul 0.8611} & {\ul 0.7152} & 0.7802 & 0.8645 & 0.7382 & 0.5172 & 0.6337 & 0.7378 \\ \hline
\textbf{DCN} & \textbf{0.8730$^{**}$} & \textbf{0.7218$^{*}$} & \textbf{0.7894$^{**}$} & \textbf{0.8704$^{*}$} & \textbf{0.7771$^{**}$} & \textbf{0.5537} & \textbf{0.6608} & \textbf{0.7731$^{**}$} \\ \hline
\end{tabular}}
\end{table}

\begin{table}[t!]
\caption{User recommendation performance comparisons on Amazon Toys, Amazon Games, Taobao, and Micro-video \chenc{(bold means $p$-value < 0.05, bold$^{*}$ means $p$-value < 0.01, and bold$^{**}$ means $p$-value < 0.001). }}\label{tbl:overall_user}
\setlength\tabcolsep{3pt}
\chenc{
\begin{tabular}{ccccccccc}
\hline
\textbf{Dataset} & \multicolumn{4}{c}{\textbf{Amazon Toys}} & \multicolumn{4}{c}{\textbf{Amazon Games}} \\ \hline
\textbf{Metric} & \textbf{AUC} & \textbf{MRR} & \textbf{NDCG} & \textbf{WAUC} & \textbf{AUC} & \textbf{MRR} & \textbf{NDCG} & \textbf{WAUC} \\ \hline
\textbf{NCF} & 0.6193 & 0.3861 & 0.5289 & 0.6130 & 0.7347 & 0.5017 & 0.6202 & 0.7296 \\ \hline
\textbf{LightGCN} & 0.6141 & 0.3555 & 0.5056 & 0.6006 & 0.6558 & 0.3885 & 0.5318 & 0.6358 \\ \hline
\textbf{DIN} & 0.6027 & 0.3777 & 0.5216 & 0.5883 & 0.7442 & 0.5005 & 0.6197 & 0.7343 \\ \hline
\textbf{Caser} & 0.6046 & 0.3895 & 0.5297 & 0.5809 & 0.7546 & 0.5382 & 0.6481 & 0.7512 \\ \hline
\textbf{GRU4REC} & 0.6424 & 0.4420 & 0.5696 & 0.6028 & {\ul 0.8015} & {\ul 0.6294} & {\ul 0.7168} & {\ul 0.7850} \\ \hline
\textbf{DIEN} & 0.5812 & 0.3570 & 0.5048 & 0.5620 & 0.7481 & 0.5363 & 0.6461 & 0.7370 \\ \hline
\textbf{SASRec} & 0.6264 & 0.4036 & 0.5416 & 0.6067 & 0.7630 & 0.5448 & 0.6531 & 0.7536 \\ \hline
\textbf{SLi-Rec} & 0.6763 & 0.4407 & 0.5720 & 0.6715 & 0.7434 & 0.5197 & 0.6338 & 0.7374 \\ \hline
\textbf{SURGE} & 0.6792 & 0.4433 & 0.5733 & 0.6724 & 0.7469 & 0.5228 & 0.6353 & 0.7385 \\ \hline
\textbf{BERT4Rec} & 0.5993 & 0.3864 & 0.5267 & 0.5662 & 0.7776 & 0.56 & 0.6651 & 0.7687 \\ \hline
\textbf{$S^3$-Rec} & 0.6077 & 0.3975 & 0.5371 & 0.5776 & 0.7891 & 0.5712 & 0.6763 & 0.7792 \\ \hline
\textbf{DuoRec} & {\ul 0.6795} & {\ul 0.4436} & {\ul 0.5755} & {\ul 0.6742} & 0.7521 & 0.5248 & 0.6389 & 0.7426 \\ \hline
\textbf{DCN} & \textbf{0.6869$^{**}$} & \textbf{0.4632$^{**}$} & \textbf{0.5893$^{**}$} & \textbf{0.6810$^{**}$} & \textbf{0.8401$^{**}$} & \textbf{0.6747$^{**}$} & \textbf{0.7520$^{**}$} & \textbf{0.8208$^{**}$} \\ \hline
\textbf{Dataset} & \multicolumn{4}{c}{\textbf{Taobao}} & \multicolumn{4}{c}{\textbf{Micro-video}} \\ \hline
\textbf{Metric} & \textbf{AUC} & \textbf{MRR} & \textbf{NDCG} & \textbf{WAUC} & \textbf{AUC} & \textbf{MRR} & \textbf{NDCG} & \textbf{WAUC} \\ \hline
\textbf{NCF} & 0.7458 & 0.5232 & 0.6367 & 0.7433 & 0.4946 & 0.2799 & 0.4443 & 0.4924 \\ \hline
\textbf{LightGCN} & 0.672 & 0.4297 & 0.5655 & 0.6966 & {\ul 0.7667} & 0.5569 & 0.6625 & {\ul 0.7564} \\ \hline
\textbf{DIN} & 0.6862 & 0.4515 & 0.5806 & 0.6784 & 0.5899 & 0.3475 & 0.4991 & 0.5861 \\ \hline
\textbf{Caser} & 0.7652 & 0.5761 & 0.6761 & 0.7520 & 0.6579 & 0.4205 & 0.5557 & 0.6424 \\ \hline
\textbf{GRU4REC} & 0.8128 & 0.6728 & {\ul 0.7497} & 0.8028 & 0.5029 & 0.3557 & 0.5001 & 0.4828 \\ \hline
\textbf{DIEN} & 0.7179 & 0.5091 & 0.6245 & 0.7089 & 0.7175 & 0.4666 & 0.5935 & 0.7133 \\ \hline
\textbf{SASRec} & 0.7829 & 0.5798 & 0.6801 & 0.7778 & 0.7632 & 0.5539 & 0.6598 & 0.7525 \\ \hline
\textbf{SLi-Rec} & 0.8264 & 0.6627 & 0.7431 & 0.8169 & 0.6256 & 0.4359 & 0.5664 & 0.6281 \\ \hline
\textbf{SURGE} & 0.8287 & 0.6648 & 0.7489 & {\ul 0.8231} & 0.6296 & 0.4394 & 0.5682 & 0.6294 \\ \hline
\textbf{BERT4Rec} & 0.7842 & 0.6063 & 0.6995 & 0.7749 & 0.7225 & 0.5534 & 0.6571 & 0.7107 \\ \hline
\textbf{$S^3$-Rec} & 0.7954 & 0.6175 & 0.7003 & 0.7861 & 0.7236 & {\ul 0.5541} & {\ul 0.6582} & 0.7121 \\ \hline
\textbf{DuoRec} & {\ul 0.8313} & {\ul 0.6679} & 0.7488 & 0.8221 & 0.6304 & 0.4412 & 0.5713 & 0.6332 \\ \hline
\textbf{DCN} & \textbf{0.8662$^{**}$} & \textbf{0.7172$^{**}$} & \textbf{0.7853$^{**}$} & \textbf{0.8584$^{**}$} & \textbf{0.7816$^{**}$} & \textbf{0.5669$^{**}$} & \textbf{0.6700$^{**}$} & \textbf{0.7652$^{**}$} \\ \hline
\end{tabular}}
\end{table}

\subsection{\chenc{Overall Performance (RQ1)}}
We show the performance of item recommendation in Table~\ref{tbl:overall_item}, from which we have the following observations.
\begin{itemize}[leftmargin=*]
\item \textbf{Our DCN achieves the best performance in user-centric perspective}. Our model achieves the best performance compared with these seven baselines for four metrics and datasets in item recommendation. Specifically, AUC is significantly~\cite{DIN} improved by 3.28\%, 2.08\%, \chenc{1.38}\%, and \chenc{3.59}\% \chenc{($p$-value < 0.05)} on the Amazon Toys, Amazon Games, Taobao, and Micro-video, respectively, when comparing DCN with other baselines. 
The improvements show that extracting dynamic item trends for auxiliary training based on contrastive learning can promote item recommendation.  

\item \chenc{\textbf{Contrastive learning can promote user-centric sequential recommendation}. DuoRec generally achieves the second best performance on the Amazon Toys, Amazon Games and Taobao datasets, while BERT4Rec and $S^3$-Rec performs decent on the Micro-video dataset. It shows the effectiveness of the contrastive learning method for promoting sequential recommendation.}

contrastive learning 
SLi-Rec is comparable with GRU4REC on two Amazon datasets with shorter item sequences but outperforms GRU4REC on the Taobao and Micro-video with longer item sequences, which illustrates its effectiveness in long-term modeling for long sequence cases. Besides, SASRec also shows its ability for extremely long sequential modeling on the Micro-video. Though existing sequential recommendation models are advanced and capable for both long and short-term modelings, it still encounters the sparse data bottleneck as our proposed approaches facilitate them into a better stage. Therefore, we tend to enhance the sequential learning by incorporating user sequence for auxiliary contrastive training instead of generating sparser augment data by dropout strategy.
\item \textbf{Sequential recommendation suffers from sparse data}. SLi-Rec is comparable with GRU4REC on two Amazon datasets with shorter item sequences but outperforms GRU4REC on the Taobao and Micro-video with longer item sequences, which illustrates its effectiveness in long-term modeling for long sequence cases. Besides, SASRec also shows its ability for extremely long sequential modeling on the Micro-video. Though existing sequential recommendation models are advanced and capable for both long and short-term modelings, it still encounters the sparse data bottleneck as our proposed approaches facilitate them into a better stage. Therefore, we tend to enhance the sequential learning by incorporating user sequence for auxiliary contrastive training instead of generating sparser augment data by dropout strategy.
\end{itemize}

\begin{table}[t!]
\caption{Ablation study on contrastive learning by user-centric evaluation with different backbones as encoders.}\label{tbl:ablation_item}
\setlength\tabcolsep{3pt}
\chenc{
\begin{tabular}{ccccccccc}
\hline
\textbf{Model   (Dataset)} & \multicolumn{4}{c}{\textbf{SLi-Rec (Amazon Toys)}} & \multicolumn{4}{c}{\textbf{GRU4REC (Amazon Games)}} \\ \hline
\textbf{Component} & \textbf{w/o RT} & \textbf{w/o DR} & \textbf{w/o DI} & \textbf{DCN} & \textbf{w/o RT} & \textbf{w/o DR} & \textbf{w/o DI} & \textbf{DCN} \\ \hline
\textbf{AUC} & 0.7452 & 0.7448 & 0.7289 & \textbf{0.7471} & 0.8146 & 0.8135 & 0.8056 & \textbf{0.8235} \\ \hline
\textbf{MRR} & 0.5098 & 0.5095 & 0.5069 & \textbf{0.5110} & 0.6530 & 0.6521 & 0.6597 & \textbf{0.6600} \\ \hline
\textbf{NDCG} & 0.6262 & 0.6259 & 0.6239 & \textbf{0.6274} & 0.7369 & 0.7362 & 0.7419 & \textbf{0.7422} \\ \hline
\textbf{WAUC} & 0.7331 & 0.7322 & 0.7290 & \textbf{0.7371} & 0.8304 & 0.8298 & 0.8341 & \textbf{0.8351} \\ \hline
\textbf{Model (Dataset)} & \multicolumn{4}{c}{\textbf{SLi-Rec (Taobao)}} & \multicolumn{4}{c}{\textbf{SASRec (Micro-video)}} \\ \hline
\textbf{Component} & \textbf{w/o RT} & \textbf{w/o DR} & \textbf{w/o DI} & \textbf{DCN} & \textbf{w/o RT} & \textbf{w/o DR} & \textbf{w/o DI} & \textbf{DCN} \\ \hline
\textbf{AUC} & 0.8722 & 0.8711 & 0.8564 & \textbf{0.8730} & 0.7542 & 0.7558 & 0.7469 & \textbf{0.7771} \\ \hline
\textbf{MRR} & 0.7216 & 0.7214 & 0.7042 & \textbf{0.7218} & 0.5413 & 0.5422 & 0.5362 & \textbf{0.5537} \\ \hline
\textbf{NDCG} & 0.7892 & 0.7889 & 0.7760 & \textbf{0.7894} & 0.6510 & 0.6517 & 0.6476 & \textbf{0.6608} \\ \hline
\textbf{WAUC} & 0.8695 & 0.8683 & 0.8616 & \textbf{0.8704} & 0.7592 & 0.7602 & 0.7558 & \textbf{0.7731} \\ \hline
\end{tabular}
}
\end{table}

\begin{table}[t!]
\caption{Ablation study on contrastive learning by item-centric evaluation with different backbones as encoders.}\label{tbl:ablation_user}
\setlength\tabcolsep{3pt}
\chenc{
\begin{tabular}{ccccccccc}
\hline
\textbf{Model   (Dataset)} & \multicolumn{4}{c}{\textbf{SLi-Rec (Amazon Toys)}} & \multicolumn{4}{c}{\textbf{GRU4REC (Amazon Games)}} \\ \hline
\textbf{Component} & \textbf{w/o RT} & \textbf{w/o DR} & \textbf{w/o DI} & \textbf{DCN} & \textbf{w/o RT} & \textbf{w/o DR} & \textbf{w/o DI} & \textbf{DCN} \\ \hline
\textbf{AUC} & 0.6807 & 0.6799 & 0.6640 & \textbf{0.6869} & 0.7964 & 0.7931 & 0.8270 & \textbf{0.8401} \\ \hline
\textbf{MRR} & 0.4559 & 0.4550 & 0.4276 & \textbf{0.4632} & 0.6273 & 0.6237 & 0.6686 & \textbf{0.6747} \\ \hline
\textbf{NDCG} & 0.5833 & 0.5825 & 0.5613 & \textbf{0.5893} & 0.7155 & 0.7127 & 0.7471 & \textbf{0.7520} \\ \hline
\textbf{WAUC} & 0.6680 & 0.6663 & 0.6487 & \textbf{0.6810} & 0.7863 & 0.7837 & 0.8103 & \textbf{0.8208} \\ \hline
\textbf{Model (Dataset)} & \multicolumn{4}{c}{\textbf{SLi-Rec (Taobao)}} & \multicolumn{4}{c}{\textbf{SASRec (Micro-video)}} \\ \hline
\textbf{Component} & \textbf{w/o RT} & \textbf{w/o DR} & \textbf{w/o DI} & \textbf{DCN} & \textbf{w/o RT} & \textbf{w/o DR} & \textbf{w/o DI} & \textbf{DCN} \\ \hline
\textbf{AUC} & 0.8632 & 0.8619 & 0.8044 & \textbf{0.8662} & 0.7424 & 0.7472 & 0.6609 & \textbf{0.7816} \\ \hline
\textbf{MRR} & 0.7169 & 0.7167 & 0.6380 & \textbf{0.7172} & 0.5114 & 0.5182 & 0.4301 & \textbf{0.5669} \\ \hline
\textbf{NDCG} & 0.7849 & 0.7847 & 0.7234 & \textbf{0.7853} & 0.6270 & 0.6323 & 0.5628 & \textbf{0.6700} \\ \hline
\textbf{WAUC} & 0.8555 & 0.8543 & 0.7882 & \textbf{0.8584} & 0.7236 & 0.7287 & 0.6393 & \textbf{0.7652} \\ \hline
\end{tabular}}
\end{table}

Besides, we also have some new insights from the item-centric perspective for the overall performance as shown in Table~\ref{tbl:overall_user} against the previous SIGIR version~\cite{dcn22}.
\begin{itemize}[leftmargin=*]
\item \textbf{Our DCN outperforms baselines sharply in item-centric perspective}. Our model achieves the best performance compared with these seven baselines for four metrics and datasets in user recommendation. Specifically, AUC is improved significantly by 1.57\%, 4.82\%, \chenc{4.20}\%, and \chenc{1.94}\%  \chenc{($p$-value < 0.001)} on Amazon Toys, Amazon Games, Taobao, and Micro-video, respectively, when comparing DCN with other baselines. 
\chenc{Unlike those on item recommendation, the improvements on user recommendation are more significant with all $p$-value < 0.001 on all metrics. }
The improvements show that the next user prediction can well model the item trends and promote user recommendations.  

\item \chenc{\textbf{Existing contrastive learning method fail to promote item-centric sequential recommendation}. Three contrastive method becomes unstable in item-centric evaluation, while they all perform poor on the Amazon Games dataset, which shows the superiority of our proposed DCN to promote the performance in not only user-centric evaluation but also item-centric evaluation.}

\item \textbf{User-centric sequential recommendation performs poor in user ranking}. Both traditional non-sequential and sequential recommendations generally perform poorly in user recommendation as they truly do not explicitly model user sequence. And some effective designs, such as long-short term modeling for item sequence of SLi-Rec, even become useless in user recommendation as it performs extremely poorly in the Micro-video dataset. Thus we introduce user sequence modeling into the traditional sequential modeling and improve existing item sequential models sharply for all user recommendation metrics on four datasets. 
\end{itemize}

\chenc{To sum up, our proposed user\&item-centric sequential recommendation framework promotes user-item interaction modeling from both user-centric and item-centric evaluations.}

\subsection{\chenc{Ablation Study (RQ2)}}

As for the further evaluations, \chenc{we first study the newly added Representation Transformation (RT) of item sequence to user representation and user sequence to item representation, respectively, in Eq.\eqref{eq:uti_space} and Eq.\eqref{eq:itu_space}}.
Besides, we also study the performances of our model without Dual Representation (DR) and Dual Interest (DI) contrastive learning of Eq.\eqref{eq:dr} and Eq.\eqref{eq:di}, respectively, at high-level. We evaluate the effectiveness of each component on four user-centric evaluation metrics under Amazon Toys, Amazon Games, Taobao, and Micro-video datasets, and the results are shown in Table~\ref{tbl:ablation_item}. It can be observed that: 

\begin{itemize}[leftmargin=*]

\item \chenc{\textbf{Representation transformation improves dual representation learning}. The performances detaching RT module generally outperform those detaching DR on the Amazon Toys, Amazon Games, and Taobao datasets, which means RT further improves DR. However, on the Micro-video dataset, the performance detaching RT is outperformed by that detaching the whole DR module, which means directly performing DR without RT may indeed harm the performance. This may be because there are more users and items on the Micro-video dataset, resulting in more complex spaces that are hard for direct contrastive learning without representation transformation. In other words, our newly added Representation Transformation can further improve and even rescues Dual Representation contrastive learning.
}

\item \textbf{Dual interest contrastive learning is more effective}. DI module is generally more effective than the DR module on four datasets except for some evaluation metrics on Amazon Games dataset, which is because the prediction objective can generate more supervised signals.
\item \textbf{Proposed components are more effective on large models with many parameters}. These two contrastive learning approaches are generally more effective on the SASRec backbone with a lot of parameters (with AUC improvement up to 4\%), which demonstrates the advantage of our self-supervised approaches for learning effective representation under sparse data.
\end{itemize}

Besides, we also study the performance of our model without Dual Representation (DR) and Dual Interest (DI) contrastive learning on four item-centric evaluation metrics under four datasets, and the results are shown in Table~\ref{tbl:ablation_user}. From the result, we can have the following observations.
\begin{itemize}[leftmargin=*]
\item \chenc{\textbf{Representation transformation improves dual representation learning}. Our newly added RT module can further improve DR module even when DT module performs poor, which is generally consistent with the observation of user-centric evaluation.}

\item  \textbf{Proposed components are more effective on item-centric evaluation} The boost effectiveness of our proposed modules is more obvious on item-centric evaluation (with AUC improvement up to 18.3\% for SASRec on Micro-video dataset) than user-centric evaluation, which illustrates there is more room for improvement on user recommendation for than item recommendation for existing next item prediction.
\item \textbf{Proposed components are more effective on large models with many parameters} These two contrastive learning approaches are generally more effective on the SASRec backbone, following that observed in the ablation result of the user-centric evaluation.
\end{itemize}

\subsection{\chenc{Study of Encoder Backbone (RQ3)}}

\begin{figure}[t]
\includegraphics[width = \linewidth]{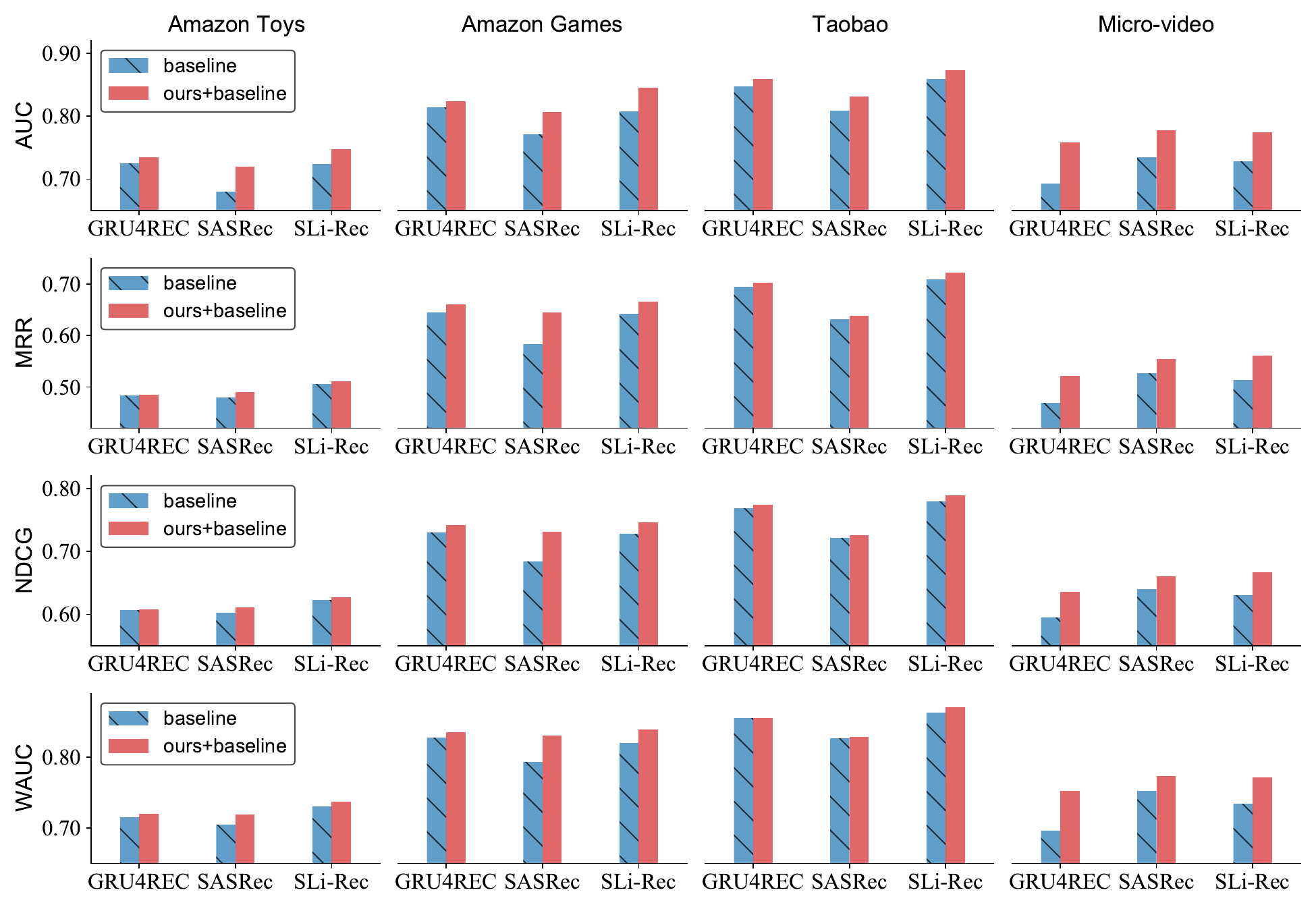}
\caption{Item recommendation performance of our proposed approaches with different backbones as encoders.}\label{fig:backbone_item}
\end{figure}
\begin{figure}[t]
\includegraphics[width = \linewidth]{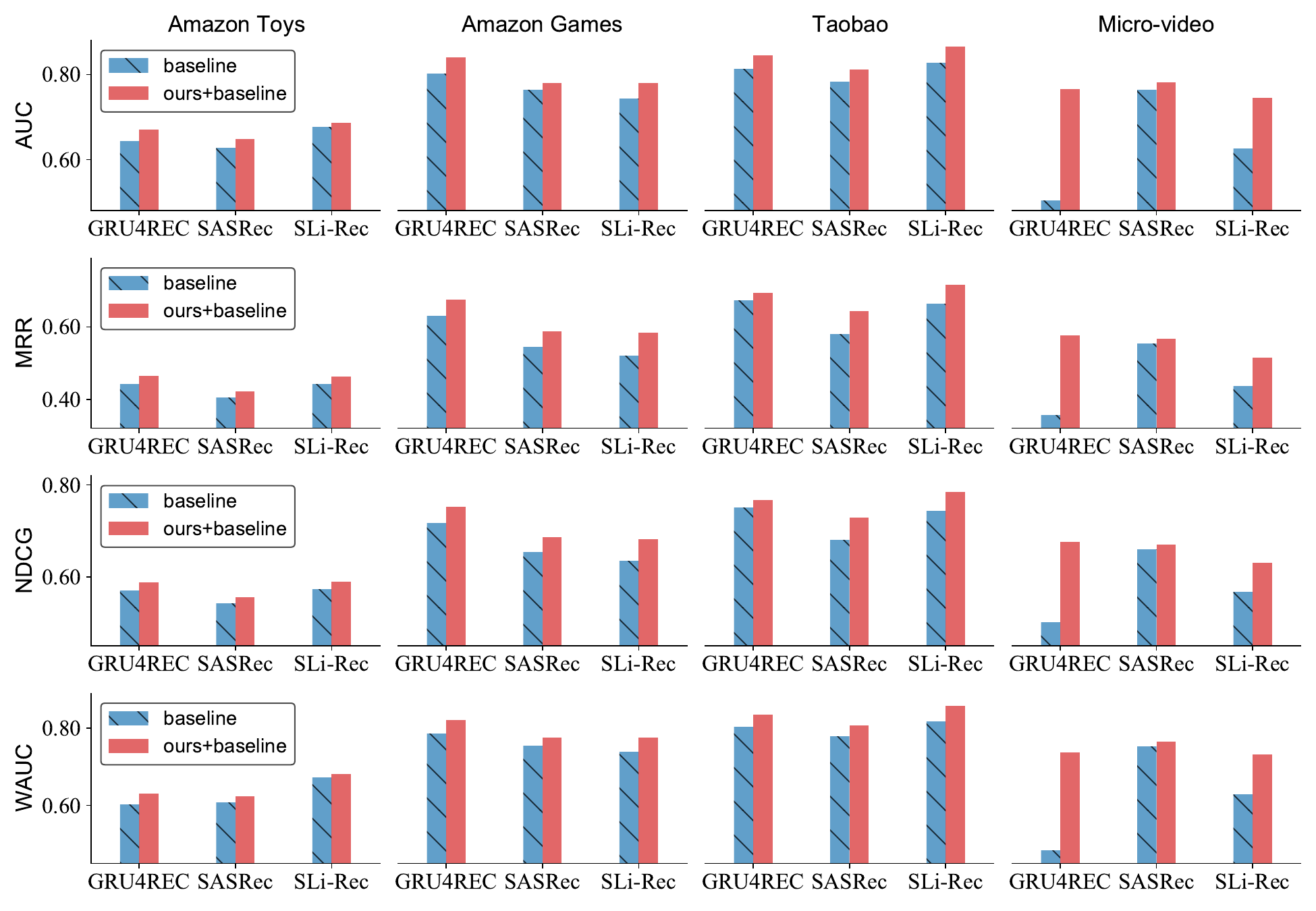}
\caption{User recommendation performance of our proposed approaches with different backbones as encoders.}\label{fig:backbone_user}
\end{figure}
In this section, we tend to verify the generalization of our proposed approaches to three models with better performance, i.e., GRU4REC, SASRec, and SLi-Rec. The user-centric evaluation results under four datasets of our approaches exploiting different backbones as encoders are as shown in Figure~\ref{fig:backbone_item}, where we can observe that:
\begin{itemize}[leftmargin=*]
\item \textbf{Proposed components are model-agnostic}. Our self-supervised approaches can boost the performance of these three backbones. The improvement is generally most obvious under SASRec with vast parameters and prone to overfitting under sparse data on e-commerce datasets such as Amazon Toys, Amazon Games, and Taobao, which verifies our approaches are more effective for sequential models requiring a lot of data for training.
\item \textbf{Contrastive learning for long-term representation is effective}. Compared with the results in the previous SIGIR version~\cite{dcn22}, improvement for SLi-Rec is promoted under SLi-Rec after introducing contrastive learning for long-term representation, which means though long and short-term modelings have provided SLi-Rec with sufficient prior knowledge our new design still can facilitate it into a better stage when it has been promoted with sufficient prior knowledge; when we achieve this improvement by performing contrastive learning on long-term representation with static embedding (i.e., target user or item embeddings), this observation also sustains our opinion in the previous version: "long term interest can be treated as static interest~\cite{dcn22}".
\end{itemize}
However, we also have additional insights as shown in Figre~\ref{fig:backbone_user} from item-centric perspective.
\begin{itemize}[leftmargin=*]
\item \textbf{Backbones are boosted more sharply on item-centric recommendation}. Our self-supervised approaches can boost the performance on item-centric evaluation more sharply than that on user-centric evaluation equipped with these three backbones, which means the introduction of user sequence can well achieve the user recommendation while promoting the item recommendation, hitting two birds with one stone.
\item \textbf{Short-term backbones are boosted more sharply on long sequence}. The improvement on SASRec becomes most minor in the Micro-video dataset when GRU4REC achieves the best improvement. This is because there is an extremely long sequence in Micro-video when SASRec and SLi-Rec are initially designed to address such a problem, and GRU4REC is restricted by its solely short-term modeling; such observation also illustrates that our approaches are not only capable of data sparsity but also suitable for the model with disadvantage on prior knowledge design.
\end{itemize}
\begin{figure*}[t]
\centering
\includegraphics[width=\linewidth]{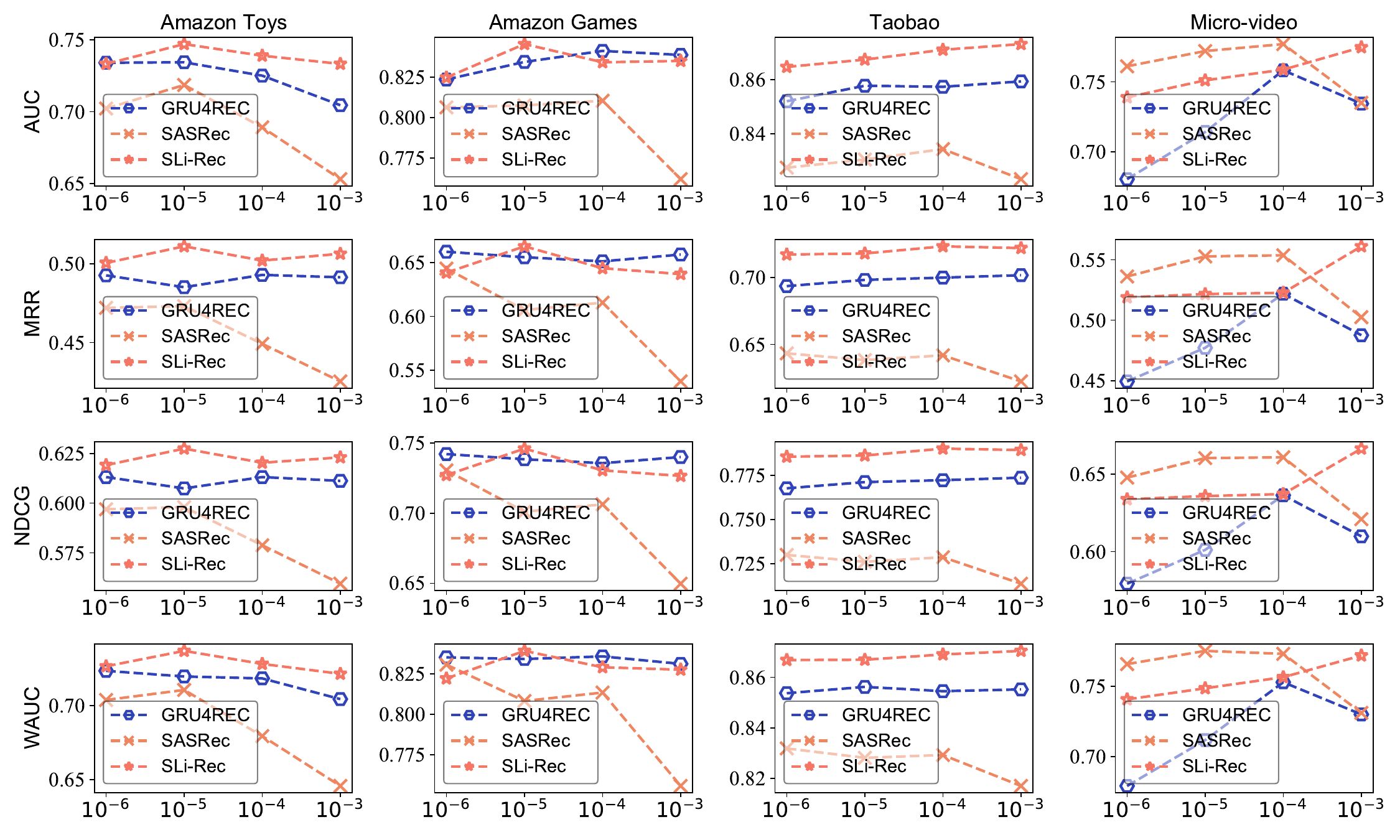}
\caption{Hyper-parameter study of regularization for contrastive learning by user-centric evaluation.}\label{fig:hyper-item}
\end{figure*}

\begin{figure*}[t]
\centering

\includegraphics[width=\linewidth]{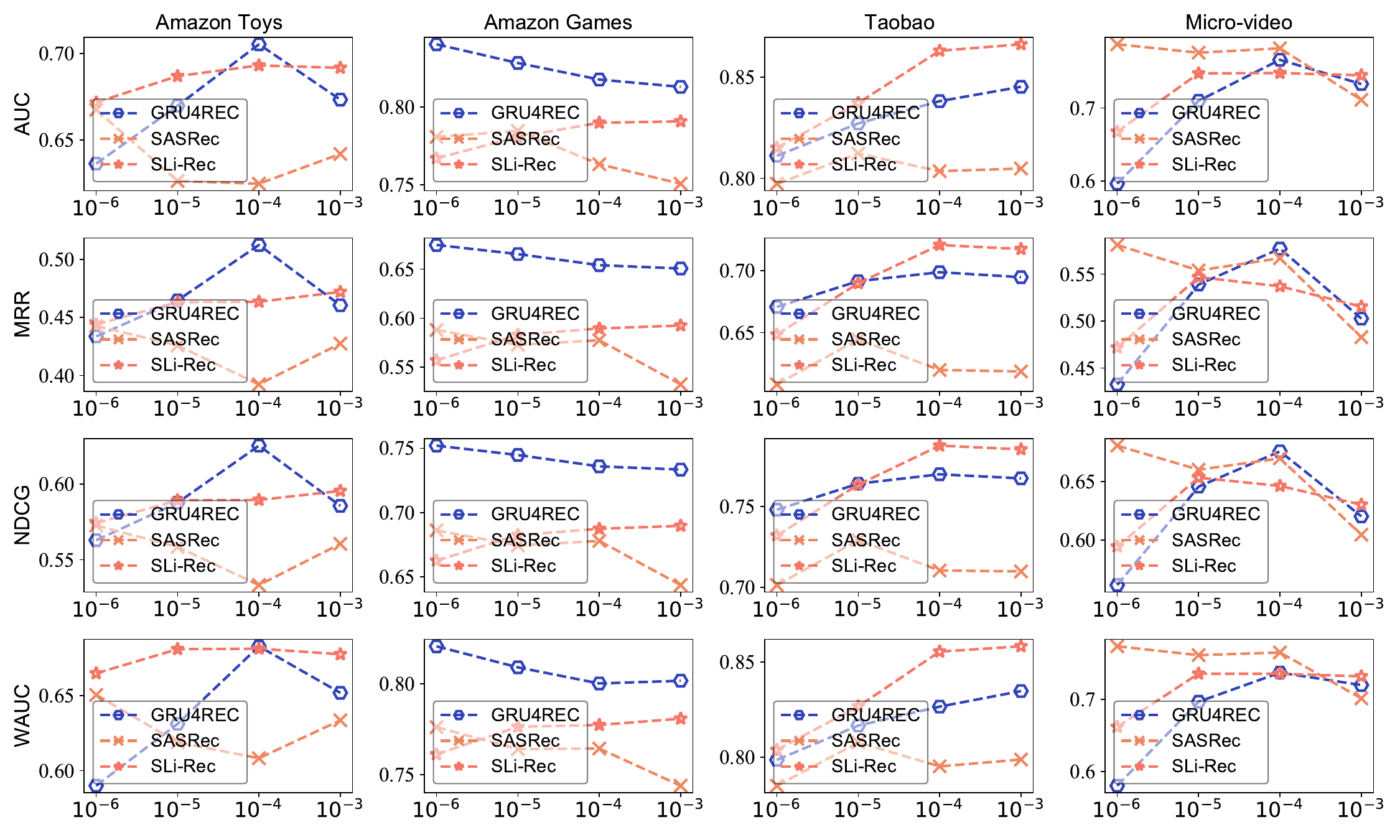}
\caption{Hyper-parameter study of regularization for contrastive learning by item-centric evaluation.}\label{fig:hyper-user}
\end{figure*}
\subsection{\chenc{Study of Hyper-Parameter (RQ4)}}~\label{sec:regularization_para}
To further understand the impact of contrastive learning, we also perform parameter searching on its regulation of it. Figure~\ref{fig:hyper-item} and \ref{fig:hyper-user} show the performance of the three best models under our framework with the regulation parameter for contrastive learning ranging from $10^{-6}$ to $10^{-3}$ in four datasets on user-centric and item-centric evaluation, where we can observe that:
\begin{itemize}[leftmargin=*]
\item \textbf{Long sequence requires stronger contrastive regularization}. Most models reach a peak when regularization parameters are at $10^{-5}$ on the Amazon Toys and Amazon Games, while they reach a peak at $10^{-3}$ on the Taobao and Micro-video with a longer sequence; this may be because it is hard for the models to capture long sequential pattern and there is often gradient vanishing problem in such case, but our proposed approaches can introduce more self-supervised signals to keep them from this dilemma to some extent.
\item \textbf{Item-centric evaluation requires stronger contrastive regularization}. The regularization parameters for best item-centric evaluation are generally greater than those for best user-centric evaluation, which is because traditional sequential recommendation has already captured the item sequential pattern to some extent; that is to say, our proposed approaches can improve the modeling of user sequence more sharply than that of item sequence.
\end{itemize}

\section{Conclusions and Future Work}

In this paper, we approached sequential recommendation from a new perspective that integrated user sequence with self-supervised learning instead of solely item sequential learning.
Existing sequential recommendation is often user-centric, which aims to predict the next interacted item for the given user with his/her historical item sequence and ignores the item-centric dimension for the item provider. 
Our introduction of item-centric modeling for user sequence not only supplemented the existing user-centric sequential recommendation with the item-centric dimension but also addressed the challenge of current self-supervised learning with sparse augmentation data by dropout strategy.
Specifically, we proposed DCN (\textbf{D}ual \textbf{C}ontrastive \textbf{N}etwork for Sequential Recommendation) with contrastive learning from representation and interest levels. Firstly, we proposed the dual representation contrastive learning, which contained user representation contrastive learning and item representation contrastive learning. Before the second kind of proposed contrastive learning approach, we propose a mixed prediction to predict the next interacted item of the given user, the next interacted user of the given item, and the collaborative filter probability without sequential modeling.   
Secondly, we proposed the dual interest contrastive learning to minimize the scaled distance between the dynamic interest for next item prediction and the static interest of matching  probability, which aimed to incorporate the static modeling into the item sequence modeling via auxiliary training. Besides, the predicted score of the next user was also self-supervised here with the next item prediction, which equipped the user-centric dimension with the item-centric dimension to further achieve mutual harmony between the user and item provider. Finally, we evaluated our proposed method on both user-centric and item-centric evaluation metrics, which rank the top items for the given user and rank the top users for the given item, respectively. Experimental results show the superiority of our proposed method against the state-of-art sequential recommendation.

As for future work, we plan to improve this paper by 1) exploring more self-supervised signals from the user sequence by item-centric dimension, such as extracting the user group representation, 2) applying the model to real-world online scenarios to further evaluate its performance, especially its exploration effect for the item provider, 3) designing specific evaluation methods to investigate the importance of considering the item-centric dimension in real-world applications.


\bibliographystyle{ACM-Reference-Format}
\bibliography{sample-base}


\begin{thebibliography}{61}


\ifx \showCODEN    \undefined \def \showCODEN     #1{\unskip}     \fi
\ifx \showDOI      \undefined \def \showDOI       #1{#1}\fi
\ifx \showISBNx    \undefined \def \showISBNx     #1{\unskip}     \fi
\ifx \showISBNxiii \undefined \def \showISBNxiii  #1{\unskip}     \fi
\ifx \showISSN     \undefined \def \showISSN      #1{\unskip}     \fi
\ifx \showLCCN     \undefined \def \showLCCN      #1{\unskip}     \fi
\ifx \shownote     \undefined \def \shownote      #1{#1}          \fi
\ifx \showarticletitle \undefined \def \showarticletitle #1{#1}   \fi
\ifx \showURL      \undefined \def \showURL       {\relax}        \fi
\providecommand\bibfield[2]{#2}
\providecommand\bibinfo[2]{#2}
\providecommand\natexlab[1]{#1}
\providecommand\showeprint[2][]{arXiv:#2}

\bibitem[\protect\citeauthoryear{Bachman, Hjelm, and Buchwalter}{Bachman
  et~al\mbox{.}}{2019}]%
        {bachman2019learning}
\bibfield{author}{\bibinfo{person}{Philip Bachman}, \bibinfo{person}{R~Devon
  Hjelm}, {and} \bibinfo{person}{William Buchwalter}.}
  \bibinfo{year}{2019}\natexlab{}.
\newblock \showarticletitle{Learning representations by maximizing mutual
  information across views}.
\newblock \bibinfo{journal}{\emph{Advances in neural information processing
  systems}}  \bibinfo{volume}{32} (\bibinfo{year}{2019}).
\newblock


\bibitem[\protect\citeauthoryear{Beutel, Covington, Jain, Xu, Li, Gatto, and
  Chi}{Beutel et~al\mbox{.}}{2018}]%
        {latentCross18}
\bibfield{author}{\bibinfo{person}{Alex Beutel}, \bibinfo{person}{Paul
  Covington}, \bibinfo{person}{Sagar Jain}, \bibinfo{person}{Can Xu},
  \bibinfo{person}{Jia Li}, \bibinfo{person}{Vince Gatto}, {and}
  \bibinfo{person}{Ed~H. Chi}.} \bibinfo{year}{2018}\natexlab{}.
\newblock \showarticletitle{Latent Cross: Making Use of Context in Recurrent
  Recommender Systems} \emph{(\bibinfo{series}{WSDM '18})}.
  \bibinfo{publisher}{Association for Computing Machinery},
  \bibinfo{address}{New York, NY, USA}, \bibinfo{pages}{46–54}.
\newblock
\showISBNx{9781450355810}
\urldef\tempurl%
\url{https://doi.org/10.1145/3159652.3159727}
\showDOI{\tempurl}


\bibitem[\protect\citeauthoryear{Chang, Gao, Zheng, Hui, Niu, Song, Jin, and
  Li}{Chang et~al\mbox{.}}{2021}]%
        {SURGE}
\bibfield{author}{\bibinfo{person}{Jianxin Chang}, \bibinfo{person}{Chen Gao},
  \bibinfo{person}{Yu Zheng}, \bibinfo{person}{Yiqun Hui},
  \bibinfo{person}{Yanan Niu}, \bibinfo{person}{Yang Song},
  \bibinfo{person}{Depeng Jin}, {and} \bibinfo{person}{Yong Li}.}
  \bibinfo{year}{2021}\natexlab{}.
\newblock \showarticletitle{Sequential Recommendation with Graph Neural
  Networks}. In \bibinfo{booktitle}{\emph{Proceedings of the 44th International
  ACM SIGIR Conference on Research and Development in Information Retrieval}}.
  \bibinfo{pages}{378--387}.
\newblock


\bibitem[\protect\citeauthoryear{Chen, Zhang, He, Nie, Liu, and Chua}{Chen
  et~al\mbox{.}}{2017}]%
        {acf17}
\bibfield{author}{\bibinfo{person}{Jingyuan Chen}, \bibinfo{person}{Hanwang
  Zhang}, \bibinfo{person}{Xiangnan He}, \bibinfo{person}{Liqiang Nie},
  \bibinfo{person}{Wei Liu}, {and} \bibinfo{person}{Tat-Seng Chua}.}
  \bibinfo{year}{2017}\natexlab{}.
\newblock \showarticletitle{Attentive Collaborative Filtering: Multimedia
  Recommendation with Item- and Component-Level Attention}. In
  \bibinfo{booktitle}{\emph{Proceedings of the 40th International ACM SIGIR
  Conference on Research and Development in Information Retrieval}} (Shinjuku,
  Tokyo, Japan) \emph{(\bibinfo{series}{SIGIR '17})}.
  \bibinfo{publisher}{Association for Computing Machinery},
  \bibinfo{address}{New York, NY, USA}, \bibinfo{pages}{335–344}.
\newblock
\showISBNx{9781450350228}
\urldef\tempurl%
\url{https://doi.org/10.1145/3077136.3080797}
\showDOI{\tempurl}


\bibitem[\protect\citeauthoryear{Chen, Beutel, Covington, Jain, Belletti, and
  Chi}{Chen et~al\mbox{.}}{2019}]%
        {chen2019top}
\bibfield{author}{\bibinfo{person}{Minmin Chen}, \bibinfo{person}{Alex Beutel},
  \bibinfo{person}{Paul Covington}, \bibinfo{person}{Sagar Jain},
  \bibinfo{person}{Francois Belletti}, {and} \bibinfo{person}{Ed~H Chi}.}
  \bibinfo{year}{2019}\natexlab{}.
\newblock \showarticletitle{Top-k off-policy correction for a REINFORCE
  recommender system}. In \bibinfo{booktitle}{\emph{Proceedings of the Twelfth
  ACM International Conference on Web Search and Data Mining}}.
  \bibinfo{pages}{456--464}.
\newblock


\bibitem[\protect\citeauthoryear{Chung, Gulcehre, Cho, and Bengio}{Chung
  et~al\mbox{.}}{2014}]%
        {GRU}
\bibfield{author}{\bibinfo{person}{Junyoung Chung}, \bibinfo{person}{Caglar
  Gulcehre}, \bibinfo{person}{Kyunghyun Cho}, {and} \bibinfo{person}{Yoshua
  Bengio}.} \bibinfo{year}{2014}\natexlab{}.
\newblock \showarticletitle{Empirical evaluation of gated recurrent neural
  networks on sequence modeling}. In \bibinfo{booktitle}{\emph{NIPS 2014
  Workshop on Deep Learning, December 2014}}.
\newblock


\bibitem[\protect\citeauthoryear{Devlin, Chang, Lee, and Toutanova}{Devlin
  et~al\mbox{.}}{2019}]%
        {devlin2018bert}
\bibfield{author}{\bibinfo{person}{Jacob Devlin}, \bibinfo{person}{Ming{-}Wei
  Chang}, \bibinfo{person}{Kenton Lee}, {and} \bibinfo{person}{Kristina
  Toutanova}.} \bibinfo{year}{2019}\natexlab{}.
\newblock \showarticletitle{{BERT:} Pre-training of Deep Bidirectional
  Transformers for Language Understanding}. In
  \bibinfo{booktitle}{\emph{Proceedings of the 2019 Conference of the North
  American Chapter of the Association for Computational Linguistics: Human
  Language Technologies, {NAACL-HLT} 2019, Minneapolis, MN, USA, June 2-7,
  2019, Volume 1 (Long and Short Papers)}},
  \bibfield{editor}{\bibinfo{person}{Jill Burstein}, \bibinfo{person}{Christy
  Doran}, {and} \bibinfo{person}{Thamar Solorio}} (Eds.).
  \bibinfo{publisher}{Association for Computational Linguistics},
  \bibinfo{pages}{4171--4186}.
\newblock
\urldef\tempurl%
\url{https://doi.org/10.18653/v1/n19-1423}
\showDOI{\tempurl}


\bibitem[\protect\citeauthoryear{Glorot and Bengio}{Glorot and Bengio}{2010}]%
        {xavier}
\bibfield{author}{\bibinfo{person}{Xavier Glorot} {and} \bibinfo{person}{Yoshua
  Bengio}.} \bibinfo{year}{2010}\natexlab{}.
\newblock \showarticletitle{Understanding the difficulty of training deep
  feedforward neural networks}. In \bibinfo{booktitle}{\emph{AISTATS}}.
  \bibinfo{pages}{249--256}.
\newblock


\bibitem[\protect\citeauthoryear{Gunawardana and Shani}{Gunawardana and
  Shani}{2015}]%
        {gunawardana_evaluating_2015}
\bibfield{author}{\bibinfo{person}{Asela Gunawardana} {and}
  \bibinfo{person}{Guy Shani}.} \bibinfo{year}{2015}\natexlab{}.
\newblock \showarticletitle{Evaluating Recommender Systems}.
\newblock In \bibinfo{booktitle}{\emph{Recommender Systems Handbook}},
  \bibfield{editor}{\bibinfo{person}{Francesco Ricci}, \bibinfo{person}{Lior
  Rokach}, {and} \bibinfo{person}{Bracha Shapira}} (Eds.).
  \bibinfo{publisher}{Springer {US}}, \bibinfo{pages}{265--308}.
\newblock
\showISBNx{978-1-4899-7637-6}
\urldef\tempurl%
\url{https://doi.org/10.1007/978-1-4899-7637-6_8}
\showDOI{\tempurl}


\bibitem[\protect\citeauthoryear{Hassani and Khasahmadi}{Hassani and
  Khasahmadi}{2020}]%
        {hassani2020contrastive}
\bibfield{author}{\bibinfo{person}{Kaveh Hassani} {and}
  \bibinfo{person}{Amir~Hosein Khasahmadi}.} \bibinfo{year}{2020}\natexlab{}.
\newblock \showarticletitle{Contrastive multi-view representation learning on
  graphs}. In \bibinfo{booktitle}{\emph{International Conference on Machine
  Learning}}. PMLR, \bibinfo{pages}{4116--4126}.
\newblock


\bibitem[\protect\citeauthoryear{He, Deng, Wang, Li, Zhang, and Wang}{He
  et~al\mbox{.}}{2020}]%
        {lightgcn}
\bibfield{author}{\bibinfo{person}{Xiangnan He}, \bibinfo{person}{Kuan Deng},
  \bibinfo{person}{Xiang Wang}, \bibinfo{person}{Yan Li},
  \bibinfo{person}{Yongdong Zhang}, {and} \bibinfo{person}{Meng Wang}.}
  \bibinfo{year}{2020}\natexlab{}.
\newblock \showarticletitle{LightGCN: Simplifying and Powering Graph
  Convolution Network for Recommendation}. In
  \bibinfo{booktitle}{\emph{SIGIR}}.
\newblock


\bibitem[\protect\citeauthoryear{He, Liao, Zhang, Nie, Hu, and Chua}{He
  et~al\mbox{.}}{2017}]%
        {NCF}
\bibfield{author}{\bibinfo{person}{Xiangnan He}, \bibinfo{person}{Lizi Liao},
  \bibinfo{person}{Hanwang Zhang}, \bibinfo{person}{Liqiang Nie},
  \bibinfo{person}{Xia Hu}, {and} \bibinfo{person}{Tat-Seng Chua}.}
  \bibinfo{year}{2017}\natexlab{}.
\newblock \showarticletitle{Neural collaborative filtering}. In
  \bibinfo{booktitle}{\emph{WWW}}. \bibinfo{pages}{173--182}.
\newblock


\bibitem[\protect\citeauthoryear{Hidasi and Karatzoglou}{Hidasi and
  Karatzoglou}{2018}]%
        {hidasi2018recurrent}
\bibfield{author}{\bibinfo{person}{Bal{\'a}zs Hidasi} {and}
  \bibinfo{person}{Alexandros Karatzoglou}.} \bibinfo{year}{2018}\natexlab{}.
\newblock \showarticletitle{Recurrent neural networks with top-k gains for
  session-based recommendations}. In \bibinfo{booktitle}{\emph{Proceedings of
  the 27th ACM international conference on information and knowledge
  management}}. \bibinfo{pages}{843--852}.
\newblock


\bibitem[\protect\citeauthoryear{Hidasi, Karatzoglou, Baltrunas, and
  Tikk}{Hidasi et~al\mbox{.}}{2016}]%
        {GRU4REC}
\bibfield{author}{\bibinfo{person}{Bal{\'a}zs Hidasi},
  \bibinfo{person}{Alexandros Karatzoglou}, \bibinfo{person}{Linas Baltrunas},
  {and} \bibinfo{person}{Domonkos Tikk}.} \bibinfo{year}{2016}\natexlab{}.
\newblock \showarticletitle{Session-based recommendations with recurrent neural
  networks}. In \bibinfo{booktitle}{\emph{ICLR}}.
\newblock


\bibitem[\protect\citeauthoryear{Hjelm, Fedorov, Lavoie-Marchildon, Grewal,
  Bachman, Trischler, and Bengio}{Hjelm et~al\mbox{.}}{2019}]%
        {hjelm2018learning}
\bibfield{author}{\bibinfo{person}{R~Devon Hjelm}, \bibinfo{person}{Alex
  Fedorov}, \bibinfo{person}{Samuel Lavoie-Marchildon}, \bibinfo{person}{Karan
  Grewal}, \bibinfo{person}{Phil Bachman}, \bibinfo{person}{Adam Trischler},
  {and} \bibinfo{person}{Yoshua Bengio}.} \bibinfo{year}{2019}\natexlab{}.
\newblock \showarticletitle{Learning deep representations by mutual information
  estimation and maximization}. In \bibinfo{booktitle}{\emph{International
  Conference on Learning Representations}}.
\newblock
\urldef\tempurl%
\url{https://openreview.net/forum?id=Bklr3j0cKX}
\showURL{%
\tempurl}


\bibitem[\protect\citeauthoryear{Hochreiter and Schmidhuber}{Hochreiter and
  Schmidhuber}{1997}]%
        {LSTM}
\bibfield{author}{\bibinfo{person}{Sepp Hochreiter} {and}
  \bibinfo{person}{J{\"u}rgen Schmidhuber}.} \bibinfo{year}{1997}\natexlab{}.
\newblock \showarticletitle{Long short-term memory}.
\newblock \bibinfo{journal}{\emph{Neural computation}} \bibinfo{volume}{9},
  \bibinfo{number}{8} (\bibinfo{year}{1997}), \bibinfo{pages}{1735--1780}.
\newblock


\bibitem[\protect\citeauthoryear{Hu, Liu, Gomes, Zitnik, Liang, Pande, and
  Leskovec}{Hu et~al\mbox{.}}{2020}]%
        {hu2019strategies}
\bibfield{author}{\bibinfo{person}{Weihua Hu}, \bibinfo{person}{Bowen Liu},
  \bibinfo{person}{Joseph Gomes}, \bibinfo{person}{Marinka Zitnik},
  \bibinfo{person}{Percy Liang}, \bibinfo{person}{Vijay Pande}, {and}
  \bibinfo{person}{Jure Leskovec}.} \bibinfo{year}{2020}\natexlab{}.
\newblock \showarticletitle{Strategies for pre-training graph neural networks}.
\newblock  (\bibinfo{year}{2020}).
\newblock


\bibitem[\protect\citeauthoryear{Kang and McAuley}{Kang and McAuley}{2018}]%
        {SASRec}
\bibfield{author}{\bibinfo{person}{Wang-Cheng Kang} {and}
  \bibinfo{person}{Julian McAuley}.} \bibinfo{year}{2018}\natexlab{}.
\newblock \showarticletitle{Self-attentive sequential recommendation}. In
  \bibinfo{booktitle}{\emph{2018 IEEE International Conference on Data Mining
  (ICDM)}}. IEEE, \bibinfo{pages}{197--206}.
\newblock


\bibitem[\protect\citeauthoryear{Kingma and Ba}{Kingma and Ba}{2015}]%
        {Adam}
\bibfield{author}{\bibinfo{person}{Diederik~P. Kingma} {and}
  \bibinfo{person}{Jimmy Ba}.} \bibinfo{year}{2015}\natexlab{}.
\newblock \showarticletitle{Adam: {A} Method for Stochastic Optimization}. In
  \bibinfo{booktitle}{\emph{{ICLR}}}.
\newblock


\bibitem[\protect\citeauthoryear{Koren, Bell, and Volinsky}{Koren
  et~al\mbox{.}}{2009}]%
        {koren2009matrix}
\bibfield{author}{\bibinfo{person}{Yehuda Koren}, \bibinfo{person}{Robert
  Bell}, {and} \bibinfo{person}{Chris Volinsky}.}
  \bibinfo{year}{2009}\natexlab{}.
\newblock \showarticletitle{Matrix factorization techniques for recommender
  systems}.
\newblock \bibinfo{journal}{\emph{Computer}} \bibinfo{volume}{42},
  \bibinfo{number}{8} (\bibinfo{year}{2009}).
\newblock


\bibitem[\protect\citeauthoryear{Krizhevsky, Sutskever, and Hinton}{Krizhevsky
  et~al\mbox{.}}{2012}]%
        {CNN}
\bibfield{author}{\bibinfo{person}{Alex Krizhevsky}, \bibinfo{person}{Ilya
  Sutskever}, {and} \bibinfo{person}{Geoffrey~E. Hinton}.}
  \bibinfo{year}{2012}\natexlab{}.
\newblock \showarticletitle{Imagenet classification with deep convolutional
  neural networks}.
\newblock   \bibinfo{volume}{25} (\bibinfo{year}{2012}),
  \bibinfo{pages}{1097--1105}.
\newblock


\bibitem[\protect\citeauthoryear{Lin, Gao, Li, Zheng, Li, Jin, and Li}{Lin
  et~al\mbox{.}}{2022a}]%
        {dcn22}
\bibfield{author}{\bibinfo{person}{Guanyu Lin}, \bibinfo{person}{Chen Gao},
  \bibinfo{person}{Yinfeng Li}, \bibinfo{person}{Yu Zheng},
  \bibinfo{person}{Zhiheng Li}, \bibinfo{person}{Depeng Jin}, {and}
  \bibinfo{person}{Yong Li}.} \bibinfo{year}{2022}\natexlab{a}.
\newblock \showarticletitle{Dual Contrastive Network for Sequential
  Recommendation}. In \bibinfo{booktitle}{\emph{Proceedings of the 45th
  International ACM SIGIR Conference on Research and Development in Information
  Retrieval}} (Madrid, Spain) \emph{(\bibinfo{series}{SIGIR '22})}.
  \bibinfo{publisher}{Association for Computing Machinery},
  \bibinfo{address}{New York, NY, USA}, \bibinfo{pages}{2686–2691}.
\newblock
\showISBNx{9781450387323}
\urldef\tempurl%
\url{https://doi.org/10.1145/3477495.3531918}
\showDOI{\tempurl}


\bibitem[\protect\citeauthoryear{Lin, Zhou, Wang, Da, Chen, and Wang}{Lin
  et~al\mbox{.}}{2022b}]%
        {sam22}
\bibfield{author}{\bibinfo{person}{Qianying Lin}, \bibinfo{person}{Wen-Ji
  Zhou}, \bibinfo{person}{Yanshi Wang}, \bibinfo{person}{Qing Da},
  \bibinfo{person}{Qing-Guo Chen}, {and} \bibinfo{person}{Bing Wang}.}
  \bibinfo{year}{2022}\natexlab{b}.
\newblock \showarticletitle{Sparse Attentive Memory Network for Click-through
  Rate Prediction with Long Sequences}. In \bibinfo{booktitle}{\emph{CIKM}}.
\newblock


\bibitem[\protect\citeauthoryear{Liu, Lin, Zhang, Xiao, and Tong}{Liu
  et~al\mbox{.}}{2019}]%
        {liu2019spiral}
\bibfield{author}{\bibinfo{person}{Dugang Liu}, \bibinfo{person}{Chen Lin},
  \bibinfo{person}{Zhilin Zhang}, \bibinfo{person}{Yanghua Xiao}, {and}
  \bibinfo{person}{Hanghang Tong}.} \bibinfo{year}{2019}\natexlab{}.
\newblock \showarticletitle{Spiral of silence in recommender systems}. In
  \bibinfo{booktitle}{\emph{Proceedings of the Twelfth ACM International
  Conference on Web Search and Data Mining}}. \bibinfo{pages}{222--230}.
\newblock


\bibitem[\protect\citeauthoryear{Ma, Zhou, Yang, Cui, Wang, and Zhu}{Ma
  et~al\mbox{.}}{2020b}]%
        {ma2020disentangled}
\bibfield{author}{\bibinfo{person}{Jianxin Ma}, \bibinfo{person}{Chang Zhou},
  \bibinfo{person}{Hongxia Yang}, \bibinfo{person}{Peng Cui},
  \bibinfo{person}{Xin Wang}, {and} \bibinfo{person}{Wenwu Zhu}.}
  \bibinfo{year}{2020}\natexlab{b}.
\newblock \showarticletitle{Disentangled Self-Supervision in Sequential
  Recommenders}. In \bibinfo{booktitle}{\emph{Proceedings of the 26th ACM
  SIGKDD International Conference on Knowledge Discovery \& Data Mining}}.
  \bibinfo{pages}{483--491}.
\newblock


\bibitem[\protect\citeauthoryear{Ma, Liu, and Deoras}{Ma et~al\mbox{.}}{2021}]%
        {ma2021bridging}
\bibfield{author}{\bibinfo{person}{Yifei Ma}, \bibinfo{person}{Ge Liu}, {and}
  \bibinfo{person}{Anoop Deoras}.} \bibinfo{year}{2021}\natexlab{}.
\newblock \showarticletitle{Bridging Recommendation and Marketing via Recurrent
  Intensity Modeling}. In \bibinfo{booktitle}{\emph{International Conference on
  Learning Representations}}.
\newblock


\bibitem[\protect\citeauthoryear{Ma, Narayanaswamy, Lin, and Ding}{Ma
  et~al\mbox{.}}{2020a}]%
        {ma2020temporal}
\bibfield{author}{\bibinfo{person}{Yifei Ma}, \bibinfo{person}{Balakrishnan
  Narayanaswamy}, \bibinfo{person}{Haibin Lin}, {and} \bibinfo{person}{Hao
  Ding}.} \bibinfo{year}{2020}\natexlab{a}.
\newblock \showarticletitle{Temporal-contextual recommendation in real-time}.
  In \bibinfo{booktitle}{\emph{Proceedings of the 26th ACM SIGKDD International
  Conference on Knowledge Discovery \& Data Mining}}.
  \bibinfo{pages}{2291--2299}.
\newblock


\bibitem[\protect\citeauthoryear{Mikolov, Karafi{\'a}t, Burget, Cernock{\`y},
  and Khudanpur}{Mikolov et~al\mbox{.}}{2010}]%
        {mikolov2010recurrent}
\bibfield{author}{\bibinfo{person}{Tomas Mikolov}, \bibinfo{person}{Martin
  Karafi{\'a}t}, \bibinfo{person}{Lukas Burget}, \bibinfo{person}{Jan
  Cernock{\`y}}, {and} \bibinfo{person}{Sanjeev Khudanpur}.}
  \bibinfo{year}{2010}\natexlab{}.
\newblock \showarticletitle{Recurrent neural network based language model.}. In
  \bibinfo{booktitle}{\emph{Interspeech}}, Vol.~\bibinfo{volume}{2}. Makuhari,
  \bibinfo{pages}{1045--1048}.
\newblock


\bibitem[\protect\citeauthoryear{Mladenov, Creager, Ben-Porat, Swersky, Zemel,
  and Boutilier}{Mladenov et~al\mbox{.}}{2020}]%
        {SocialWelfare}
\bibfield{author}{\bibinfo{person}{Martin Mladenov}, \bibinfo{person}{Elliot
  Creager}, \bibinfo{person}{Omer Ben-Porat}, \bibinfo{person}{Kevin Swersky},
  \bibinfo{person}{Richard Zemel}, {and} \bibinfo{person}{Craig Boutilier}.}
  \bibinfo{year}{2020}\natexlab{}.
\newblock \showarticletitle{Optimizing Long-Term Social Welfare in Recommender
  Systems: A Constrained Matching Approach}. In
  \bibinfo{booktitle}{\emph{Proceedings of the 37th International Conference on
  Machine Learning}} \emph{(\bibinfo{series}{ICML'20})}.
  \bibinfo{publisher}{JMLR.org}, Article \bibinfo{articleno}{648},
  \bibinfo{numpages}{12}~pages.
\newblock


\bibitem[\protect\citeauthoryear{Pi, Zhou, Zhang, Wang, Ren, Fan, Zhu, and
  Gai}{Pi et~al\mbox{.}}{2020}]%
        {SIM20}
\bibfield{author}{\bibinfo{person}{Qi Pi}, \bibinfo{person}{Guorui Zhou},
  \bibinfo{person}{Yujing Zhang}, \bibinfo{person}{Zhe Wang},
  \bibinfo{person}{Lejian Ren}, \bibinfo{person}{Ying Fan},
  \bibinfo{person}{Xiaoqiang Zhu}, {and} \bibinfo{person}{Kun Gai}.}
  \bibinfo{year}{2020}\natexlab{}.
\newblock \showarticletitle{Search-Based User Interest Modeling with Lifelong
  Sequential Behavior Data for Click-Through Rate Prediction}. In
  \bibinfo{booktitle}{\emph{Proceedings of the 29th ACM International
  Conference on Information and Knowledge Management}} (Virtual Event, Ireland)
  \emph{(\bibinfo{series}{CIKM '20})}. \bibinfo{publisher}{Association for
  Computing Machinery}, \bibinfo{address}{New York, NY, USA},
  \bibinfo{pages}{2685–2692}.
\newblock
\showISBNx{9781450368599}
\urldef\tempurl%
\url{https://doi.org/10.1145/3340531.3412744}
\showDOI{\tempurl}


\bibitem[\protect\citeauthoryear{Qiu, Chen, Dong, Zhang, Yang, Ding, Wang, and
  Tang}{Qiu et~al\mbox{.}}{2020}]%
        {qiu2020gcc}
\bibfield{author}{\bibinfo{person}{Jiezhong Qiu}, \bibinfo{person}{Qibin Chen},
  \bibinfo{person}{Yuxiao Dong}, \bibinfo{person}{Jing Zhang},
  \bibinfo{person}{Hongxia Yang}, \bibinfo{person}{Ming Ding},
  \bibinfo{person}{Kuansan Wang}, {and} \bibinfo{person}{Jie Tang}.}
  \bibinfo{year}{2020}\natexlab{}.
\newblock \showarticletitle{Gcc: Graph contrastive coding for graph neural
  network pre-training}. In \bibinfo{booktitle}{\emph{Proceedings of the 26th
  ACM SIGKDD International Conference on Knowledge Discovery \& Data Mining}}.
  \bibinfo{pages}{1150--1160}.
\newblock


\bibitem[\protect\citeauthoryear{Qiu, Huang, Yin, and Wang}{Qiu
  et~al\mbox{.}}{2022}]%
        {duorec}
\bibfield{author}{\bibinfo{person}{Ruihong Qiu}, \bibinfo{person}{Zi Huang},
  \bibinfo{person}{Hongzhi Yin}, {and} \bibinfo{person}{Zijian Wang}.}
  \bibinfo{year}{2022}\natexlab{}.
\newblock \showarticletitle{Contrastive learning for representation
  degeneration problem in sequential recommendation}. In
  \bibinfo{booktitle}{\emph{Proceedings of the fifteenth ACM international
  conference on web search and data mining}}. \bibinfo{pages}{813--823}.
\newblock


\bibitem[\protect\citeauthoryear{Rendle, Freudenthaler, and
  Schmidt-Thieme}{Rendle et~al\mbox{.}}{2010}]%
        {rendle2010factorizing}
\bibfield{author}{\bibinfo{person}{Steffen Rendle}, \bibinfo{person}{Christoph
  Freudenthaler}, {and} \bibinfo{person}{Lars Schmidt-Thieme}.}
  \bibinfo{year}{2010}\natexlab{}.
\newblock \showarticletitle{Factorizing personalized markov chains for
  next-basket recommendation}. In \bibinfo{booktitle}{\emph{WWW}}.
  \bibinfo{pages}{811--820}.
\newblock


\bibitem[\protect\citeauthoryear{Schnabel, Swaminathan, Singh, Chandak, and
  Joachims}{Schnabel et~al\mbox{.}}{2016}]%
        {schnabel2016recommendations}
\bibfield{author}{\bibinfo{person}{Tobias Schnabel}, \bibinfo{person}{Adith
  Swaminathan}, \bibinfo{person}{Ashudeep Singh}, \bibinfo{person}{Navin
  Chandak}, {and} \bibinfo{person}{Thorsten Joachims}.}
  \bibinfo{year}{2016}\natexlab{}.
\newblock \showarticletitle{Recommendations as treatments: Debiasing learning
  and evaluation}. In \bibinfo{booktitle}{\emph{international conference on
  machine learning}}. PMLR, \bibinfo{pages}{1670--1679}.
\newblock


\bibitem[\protect\citeauthoryear{Singh and Joachims}{Singh and
  Joachims}{2018}]%
        {Fairness}
\bibfield{author}{\bibinfo{person}{Ashudeep Singh} {and}
  \bibinfo{person}{Thorsten Joachims}.} \bibinfo{year}{2018}\natexlab{}.
\newblock \showarticletitle{Fairness of Exposure in Rankings}
  \emph{(\bibinfo{series}{KDD '18})}. \bibinfo{publisher}{Association for
  Computing Machinery}, \bibinfo{address}{New York, NY, USA},
  \bibinfo{pages}{2219–2228}.
\newblock
\showISBNx{9781450355520}
\urldef\tempurl%
\url{https://doi.org/10.1145/3219819.3220088}
\showDOI{\tempurl}


\bibitem[\protect\citeauthoryear{Su, Bayoumi, and Joachims}{Su
  et~al\mbox{.}}{2022}]%
        {MatchingMarkets}
\bibfield{author}{\bibinfo{person}{Yi Su}, \bibinfo{person}{Magd Bayoumi},
  {and} \bibinfo{person}{Thorsten Joachims}.} \bibinfo{year}{2022}\natexlab{}.
\newblock \showarticletitle{Optimizing Rankings for Recommendation in Matching
  Markets}. In \bibinfo{booktitle}{\emph{Proceedings of the ACM Web Conference
  2022}} (Virtual Event, Lyon, France) \emph{(\bibinfo{series}{WWW '22})}.
  \bibinfo{publisher}{Association for Computing Machinery},
  \bibinfo{address}{New York, NY, USA}, \bibinfo{pages}{328–338}.
\newblock
\showISBNx{9781450390965}
\urldef\tempurl%
\url{https://doi.org/10.1145/3485447.3511961}
\showDOI{\tempurl}


\bibitem[\protect\citeauthoryear{Sun, Liu, Wu, Pei, Lin, Ou, and Jiang}{Sun
  et~al\mbox{.}}{2019}]%
        {sun2019bert4rec}
\bibfield{author}{\bibinfo{person}{Fei Sun}, \bibinfo{person}{Jun Liu},
  \bibinfo{person}{Jian Wu}, \bibinfo{person}{Changhua Pei},
  \bibinfo{person}{Xiao Lin}, \bibinfo{person}{Wenwu Ou}, {and}
  \bibinfo{person}{Peng Jiang}.} \bibinfo{year}{2019}\natexlab{}.
\newblock \showarticletitle{BERT4Rec: Sequential recommendation with
  bidirectional encoder representations from transformer}. In
  \bibinfo{booktitle}{\emph{Proceedings of the 28th ACM International
  Conference on Information and Knowledge Management}}.
  \bibinfo{pages}{1441--1450}.
\newblock


\bibitem[\protect\citeauthoryear{Sun, Hoffmann, Verma, and Tang}{Sun
  et~al\mbox{.}}{2020}]%
        {sun2019infograph}
\bibfield{author}{\bibinfo{person}{Fan-Yun Sun}, \bibinfo{person}{Jordan
  Hoffmann}, \bibinfo{person}{Vikas Verma}, {and} \bibinfo{person}{Jian Tang}.}
  \bibinfo{year}{2020}\natexlab{}.
\newblock \showarticletitle{Infograph: Unsupervised and semi-supervised
  graph-level representation learning via mutual information maximization}.
\newblock  (\bibinfo{year}{2020}).
\newblock


\bibitem[\protect\citeauthoryear{Tang, Hu, and Liu}{Tang et~al\mbox{.}}{2013}]%
        {tang2013social}
\bibfield{author}{\bibinfo{person}{Jiliang Tang}, \bibinfo{person}{Xia Hu},
  {and} \bibinfo{person}{Huan Liu}.} \bibinfo{year}{2013}\natexlab{}.
\newblock \showarticletitle{Social recommendation: a review}.
\newblock \bibinfo{journal}{\emph{Social Network Analysis and Mining}}
  \bibinfo{volume}{3}, \bibinfo{number}{4} (\bibinfo{year}{2013}),
  \bibinfo{pages}{1113--1133}.
\newblock


\bibitem[\protect\citeauthoryear{Tang and Wang}{Tang and Wang}{2018}]%
        {Caser}
\bibfield{author}{\bibinfo{person}{Jiaxi Tang} {and} \bibinfo{person}{Ke
  Wang}.} \bibinfo{year}{2018}\natexlab{}.
\newblock \showarticletitle{Personalized top-n sequential recommendation via
  convolutional sequence embedding}. In \bibinfo{booktitle}{\emph{WWW}}.
  \bibinfo{pages}{565--573}.
\newblock


\bibitem[\protect\citeauthoryear{Vaswani, Shazeer, Parmar, Uszkoreit, Jones,
  Gomez, Kaiser, and Polosukhin}{Vaswani et~al\mbox{.}}{2017}]%
        {vaswani2017attention}
\bibfield{author}{\bibinfo{person}{Ashish Vaswani}, \bibinfo{person}{Noam
  Shazeer}, \bibinfo{person}{Niki Parmar}, \bibinfo{person}{Jakob Uszkoreit},
  \bibinfo{person}{Llion Jones}, \bibinfo{person}{Aidan~N Gomez},
  \bibinfo{person}{{\L}ukasz Kaiser}, {and} \bibinfo{person}{Illia
  Polosukhin}.} \bibinfo{year}{2017}\natexlab{}.
\newblock \showarticletitle{Attention is all you need}. In
  \bibinfo{booktitle}{\emph{NeurIPS}}. \bibinfo{pages}{5998--6008}.
\newblock


\bibitem[\protect\citeauthoryear{Velickovic, Fedus, Hamilton, Li{\`o}, Bengio,
  and Hjelm}{Velickovic et~al\mbox{.}}{2019}]%
        {velickovic2019deep}
\bibfield{author}{\bibinfo{person}{Petar Velickovic}, \bibinfo{person}{William
  Fedus}, \bibinfo{person}{William~L Hamilton}, \bibinfo{person}{Pietro
  Li{\`o}}, \bibinfo{person}{Yoshua Bengio}, {and} \bibinfo{person}{R~Devon
  Hjelm}.} \bibinfo{year}{2019}\natexlab{}.
\newblock \showarticletitle{Deep Graph Infomax}.
\newblock \bibinfo{journal}{\emph{ICLR (Poster)}} \bibinfo{volume}{2},
  \bibinfo{number}{3} (\bibinfo{year}{2019}), \bibinfo{pages}{4}.
\newblock


\bibitem[\protect\citeauthoryear{Wang, Zhang, Yuan, et~al\mbox{.}}{Wang
  et~al\mbox{.}}{2017}]%
        {RTB}
\bibfield{author}{\bibinfo{person}{Jun Wang}, \bibinfo{person}{Weinan Zhang},
  \bibinfo{person}{Shuai Yuan}, {et~al\mbox{.}}}
  \bibinfo{year}{2017}\natexlab{}.
\newblock \showarticletitle{Display advertising with real-time bidding (RTB)
  and behavioural targeting}.
\newblock \bibinfo{journal}{\emph{Foundations and Trends{\textregistered} in
  Information Retrieval}} \bibinfo{volume}{11}, \bibinfo{number}{4-5}
  (\bibinfo{year}{2017}), \bibinfo{pages}{297--435}.
\newblock


\bibitem[\protect\citeauthoryear{Wang, Hu, Wang, Cao, Sheng, and Orgun}{Wang
  et~al\mbox{.}}{2019a}]%
        {SRs}
\bibfield{author}{\bibinfo{person}{Shoujin Wang}, \bibinfo{person}{Liang Hu},
  \bibinfo{person}{Yan Wang}, \bibinfo{person}{Longbing Cao},
  \bibinfo{person}{Quan~Z. Sheng}, {and} \bibinfo{person}{Mehmet Orgun}.}
  \bibinfo{year}{2019}\natexlab{a}.
\newblock \showarticletitle{Sequential recommender systems: challenges,
  progress and prospects}.
\newblock  (\bibinfo{year}{2019}).
\newblock


\bibitem[\protect\citeauthoryear{Wang, Hu, Wang, Cao, Sheng, and Orgun}{Wang
  et~al\mbox{.}}{2019b}]%
        {wang_sequential_2019}
\bibfield{author}{\bibinfo{person}{Shoujin Wang}, \bibinfo{person}{Liang Hu},
  \bibinfo{person}{Yan Wang}, \bibinfo{person}{Longbing Cao},
  \bibinfo{person}{Quan~Z. Sheng}, {and} \bibinfo{person}{Mehmet Orgun}.}
  \bibinfo{year}{2019}\natexlab{b}.
\newblock \showarticletitle{Sequential recommender systems: challenges,
  progress and prospects}.
\newblock  (\bibinfo{year}{2019}).
\newblock


\bibitem[\protect\citeauthoryear{Wang, Zhang, Xu, Chen, Zhang, Zhao, and
  Wen}{Wang et~al\mbox{.}}{2021}]%
        {wang_counterfactual_2021}
\bibfield{author}{\bibinfo{person}{Zhenlei Wang}, \bibinfo{person}{Jingsen
  Zhang}, \bibinfo{person}{Hongteng Xu}, \bibinfo{person}{Xu Chen},
  \bibinfo{person}{Yongfeng Zhang}, \bibinfo{person}{Wayne~Xin Zhao}, {and}
  \bibinfo{person}{Ji-Rong Wen}.} \bibinfo{year}{2021}\natexlab{}.
\newblock \showarticletitle{Counterfactual data-augmented sequential
  recommendation}. In \bibinfo{booktitle}{\emph{Proceedings of the 44th
  International {ACM} {SIGIR} Conference on Research and Development in
  Information Retrieval}}. \bibinfo{pages}{347--356}.
\newblock


\bibitem[\protect\citeauthoryear{Wu, Lin, Gao, Tan, and Li}{Wu
  et~al\mbox{.}}{2021}]%
        {wu2021self}
\bibfield{author}{\bibinfo{person}{Lirong Wu}, \bibinfo{person}{Haitao Lin},
  \bibinfo{person}{Zhangyang Gao}, \bibinfo{person}{Cheng Tan}, {and}
  \bibinfo{person}{Stan~Z Li}.} \bibinfo{year}{2021}\natexlab{}.
\newblock \showarticletitle{Self-supervised on graphs: Contrastive, generative,
  or predictive}.
\newblock  (\bibinfo{year}{2021}).
\newblock


\bibitem[\protect\citeauthoryear{Xia, Yin, Yu, Shao, and Cui}{Xia
  et~al\mbox{.}}{2021}]%
        {xia2021self}
\bibfield{author}{\bibinfo{person}{Xin Xia}, \bibinfo{person}{Hongzhi Yin},
  \bibinfo{person}{Junliang Yu}, \bibinfo{person}{Yingxia Shao}, {and}
  \bibinfo{person}{Lizhen Cui}.} \bibinfo{year}{2021}\natexlab{}.
\newblock \showarticletitle{Self-Supervised Graph Co-Training for Session-based
  Recommendation}. In \bibinfo{booktitle}{\emph{Proceedings of the 30th ACM
  International Conference on Information \& Knowledge Management}}.
  \bibinfo{pages}{2180--2190}.
\newblock


\bibitem[\protect\citeauthoryear{Xie, Sun, Liu, Gao, Ding, and Cui}{Xie
  et~al\mbox{.}}{2020}]%
        {xie2020contrastive}
\bibfield{author}{\bibinfo{person}{Xu Xie}, \bibinfo{person}{Fei Sun},
  \bibinfo{person}{Zhaoyang Liu}, \bibinfo{person}{Jinyang Gao},
  \bibinfo{person}{Bolin Ding}, {and} \bibinfo{person}{Bin Cui}.}
  \bibinfo{year}{2020}\natexlab{}.
\newblock \showarticletitle{Contrastive pre-training for sequential
  recommendation}.
\newblock \bibinfo{journal}{\emph{WWW}} (\bibinfo{year}{2020}).
\newblock


\bibitem[\protect\citeauthoryear{Xin, Karatzoglou, Arapakis, and Jose}{Xin
  et~al\mbox{.}}{2020}]%
        {xin2020self}
\bibfield{author}{\bibinfo{person}{Xin Xin}, \bibinfo{person}{Alexandros
  Karatzoglou}, \bibinfo{person}{Ioannis Arapakis}, {and}
  \bibinfo{person}{Joemon~M Jose}.} \bibinfo{year}{2020}\natexlab{}.
\newblock \showarticletitle{Self-supervised reinforcement learning for
  recommender systems}. In \bibinfo{booktitle}{\emph{Proceedings of the 43rd
  International ACM SIGIR conference on research and development in Information
  Retrieval}}. \bibinfo{pages}{931--940}.
\newblock


\bibitem[\protect\citeauthoryear{Yang, Cui, Xuan, Wang, Belongie, and
  Estrin}{Yang et~al\mbox{.}}{2018}]%
        {yang2018unbiased}
\bibfield{author}{\bibinfo{person}{Longqi Yang}, \bibinfo{person}{Yin Cui},
  \bibinfo{person}{Yuan Xuan}, \bibinfo{person}{Chenyang Wang},
  \bibinfo{person}{Serge Belongie}, {and} \bibinfo{person}{Deborah Estrin}.}
  \bibinfo{year}{2018}\natexlab{}.
\newblock \showarticletitle{Unbiased offline recommender evaluation for
  missing-not-at-random implicit feedback}. In
  \bibinfo{booktitle}{\emph{Proceedings of the 12th ACM conference on
  recommender systems}}. \bibinfo{pages}{279--287}.
\newblock


\bibitem[\protect\citeauthoryear{Yao, Yi, Cheng, Felix, Menon, Hong, Chi, Tjoa,
  Kang, and Ettinger}{Yao et~al\mbox{.}}{2020}]%
        {yao2020self}
\bibfield{author}{\bibinfo{person}{Tiansheng Yao}, \bibinfo{person}{Xinyang
  Yi}, \bibinfo{person}{Derek~Zhiyuan Cheng}, \bibinfo{person}{X~Yu Felix},
  \bibinfo{person}{Aditya~Krishna Menon}, \bibinfo{person}{Lichan Hong},
  \bibinfo{person}{Ed~H Chi}, \bibinfo{person}{Steve Tjoa},
  \bibinfo{person}{Jieqi Kang}, {and} \bibinfo{person}{Evan Ettinger}.}
  \bibinfo{year}{2020}\natexlab{}.
\newblock \showarticletitle{Self-supervised learning for deep models in
  recommendations}.
\newblock \bibinfo{journal}{\emph{CORR}} (\bibinfo{year}{2020}).
\newblock


\bibitem[\protect\citeauthoryear{Yu, Yin, Gao, Xia, Zhang, and Viet~Hung}{Yu
  et~al\mbox{.}}{2021a}]%
        {yu2021socially}
\bibfield{author}{\bibinfo{person}{Junliang Yu}, \bibinfo{person}{Hongzhi Yin},
  \bibinfo{person}{Min Gao}, \bibinfo{person}{Xin Xia},
  \bibinfo{person}{Xiangliang Zhang}, {and} \bibinfo{person}{Nguyen~Quoc
  Viet~Hung}.} \bibinfo{year}{2021}\natexlab{a}.
\newblock \showarticletitle{Socially-aware self-supervised tri-training for
  recommendation}. In \bibinfo{booktitle}{\emph{Proceedings of the 27th ACM
  SIGKDD Conference on Knowledge Discovery \& Data Mining}}.
  \bibinfo{pages}{2084--2092}.
\newblock


\bibitem[\protect\citeauthoryear{Yu, Yin, Li, Wang, Hung, and Zhang}{Yu
  et~al\mbox{.}}{2021b}]%
        {yu2021self}
\bibfield{author}{\bibinfo{person}{Junliang Yu}, \bibinfo{person}{Hongzhi Yin},
  \bibinfo{person}{Jundong Li}, \bibinfo{person}{Qinyong Wang},
  \bibinfo{person}{Nguyen Quoc~Viet Hung}, {and} \bibinfo{person}{Xiangliang
  Zhang}.} \bibinfo{year}{2021}\natexlab{b}.
\newblock \showarticletitle{Self-supervised multi-channel hypergraph
  convolutional network for social recommendation}. In
  \bibinfo{booktitle}{\emph{Proceedings of the Web Conference 2021}}.
  \bibinfo{pages}{413--424}.
\newblock


\bibitem[\protect\citeauthoryear{Yu, Lian, Mahmoody, Liu, and Xie}{Yu
  et~al\mbox{.}}{2019}]%
        {SLIREC}
\bibfield{author}{\bibinfo{person}{Zeping Yu}, \bibinfo{person}{Jianxun Lian},
  \bibinfo{person}{Ahmad Mahmoody}, \bibinfo{person}{Gongshen Liu}, {and}
  \bibinfo{person}{Xing Xie}.} \bibinfo{year}{2019}\natexlab{}.
\newblock \showarticletitle{Adaptive User Modeling with Long and Short-Term
  Preferences for Personalized Recommendation}. In
  \bibinfo{booktitle}{\emph{IJCAI}}. \bibinfo{pages}{4213--4219}.
\newblock


\bibitem[\protect\citeauthoryear{Yuan, Wang, Li, and Qin}{Yuan
  et~al\mbox{.}}{2014}]%
        {yuan2014survey}
\bibfield{author}{\bibinfo{person}{Yong Yuan}, \bibinfo{person}{Feiyue Wang},
  \bibinfo{person}{Juanjuan Li}, {and} \bibinfo{person}{Rui Qin}.}
  \bibinfo{year}{2014}\natexlab{}.
\newblock \showarticletitle{A survey on real time bidding advertising}. In
  \bibinfo{booktitle}{\emph{Proceedings of 2014 IEEE International Conference
  on Service Operations and Logistics, and Informatics}}. IEEE,
  \bibinfo{pages}{418--423}.
\newblock


\bibitem[\protect\citeauthoryear{Zhao, Wang, Ye, Gao, Yang, and Chen}{Zhao
  et~al\mbox{.}}{2018}]%
        {zhao2018plastic}
\bibfield{author}{\bibinfo{person}{Wei Zhao}, \bibinfo{person}{Benyou Wang},
  \bibinfo{person}{Jianbo Ye}, \bibinfo{person}{Yongqiang Gao},
  \bibinfo{person}{Min Yang}, {and} \bibinfo{person}{Xiaojun Chen}.}
  \bibinfo{year}{2018}\natexlab{}.
\newblock \showarticletitle{PLASTIC: Prioritize Long and Short-term Information
  in Top-n Recommendation using Adversarial Training}. In
  \bibinfo{booktitle}{\emph{IJCAI}}. \bibinfo{pages}{3676--3682}.
\newblock


\bibitem[\protect\citeauthoryear{Zheng, Gao, Chang, Niu, Song, Jin, and
  Yong}{Zheng et~al\mbox{.}}{2022}]%
        {zheng2022disentangling}
\bibfield{author}{\bibinfo{person}{Yu Zheng}, \bibinfo{person}{Chen Gao},
  \bibinfo{person}{Jianxin Chang}, \bibinfo{person}{Yanan Niu},
  \bibinfo{person}{Yang Song}, \bibinfo{person}{Depeng Jin}, {and}
  \bibinfo{person}{Li Yong}.} \bibinfo{year}{2022}\natexlab{}.
\newblock \showarticletitle{Disentangling Long and Short-Term Interests for
  Recommendation}. In \bibinfo{booktitle}{\emph{TheWebConf}}.
\newblock


\bibitem[\protect\citeauthoryear{Zhou, Mou, Fan, Pi, Bian, Zhou, Zhu, and
  Gai}{Zhou et~al\mbox{.}}{2019}]%
        {DIEN}
\bibfield{author}{\bibinfo{person}{Guorui Zhou}, \bibinfo{person}{Na Mou},
  \bibinfo{person}{Ying Fan}, \bibinfo{person}{Qi Pi}, \bibinfo{person}{Weijie
  Bian}, \bibinfo{person}{Chang Zhou}, \bibinfo{person}{Xiaoqiang Zhu}, {and}
  \bibinfo{person}{Kun Gai}.} \bibinfo{year}{2019}\natexlab{}.
\newblock \showarticletitle{Deep interest evolution network for click-through
  rate prediction}. In \bibinfo{booktitle}{\emph{AAAI}}.
  \bibinfo{pages}{5941--5948}.
\newblock


\bibitem[\protect\citeauthoryear{Zhou, Zhu, Song, Fan, Zhu, Ma, Yan, Jin, Li,
  and Gai}{Zhou et~al\mbox{.}}{2018}]%
        {DIN}
\bibfield{author}{\bibinfo{person}{Guorui Zhou}, \bibinfo{person}{Xiaoqiang
  Zhu}, \bibinfo{person}{Chenru Song}, \bibinfo{person}{Ying Fan},
  \bibinfo{person}{Han Zhu}, \bibinfo{person}{Xiao Ma},
  \bibinfo{person}{Yanghui Yan}, \bibinfo{person}{Junqi Jin},
  \bibinfo{person}{Han Li}, {and} \bibinfo{person}{Kun Gai}.}
  \bibinfo{year}{2018}\natexlab{}.
\newblock \showarticletitle{Deep interest network for click-through rate
  prediction}. In \bibinfo{booktitle}{\emph{KDD}}. \bibinfo{pages}{1059--1068}.
\newblock


\bibitem[\protect\citeauthoryear{Zhou, Wang, Zhao, Zhu, Wang, Zhang, Wang, and
  Wen}{Zhou et~al\mbox{.}}{2020}]%
        {zhou2020s3}
\bibfield{author}{\bibinfo{person}{Kun Zhou}, \bibinfo{person}{Hui Wang},
  \bibinfo{person}{Wayne~Xin Zhao}, \bibinfo{person}{Yutao Zhu},
  \bibinfo{person}{Sirui Wang}, \bibinfo{person}{Fuzheng Zhang},
  \bibinfo{person}{Zhongyuan Wang}, {and} \bibinfo{person}{Ji-Rong Wen}.}
  \bibinfo{year}{2020}\natexlab{}.
\newblock \showarticletitle{$\text{S}^3$-Rec: Self-supervised learning for
  sequential recommendation with mutual information maximization}. In
  \bibinfo{booktitle}{\emph{Proceedings of the 29th ACM International
  Conference on Information \& Knowledge Management}}.
  \bibinfo{pages}{1893--1902}.
\newblock


\end{thebibliography}
\end{document}